\documentclass[10pt,journal]{IEEEtran}

\IEEEoverridecommandlockouts

\usepackage{amsmath,amsthm}
\usepackage{epsfig}
\usepackage{graphicx}
\usepackage{float}
\usepackage{stmaryrd}
\usepackage{cite}
\usepackage[hyphens]{url}
\usepackage[colorlinks=true,hyperindex,breaklinks]{hyperref}
\usepackage[table,hyperref,x11names]{xcolor}
\usepackage{comment}
\usepackage{array}
\usepackage{hhline}
\usepackage[ruled,vlined,linesnumbered]{algorithm2e}
\usepackage{diagbox}
\usepackage{enumitem}
\usepackage{amssymb}
\usepackage{mathrsfs}
\usepackage{multirow}
\usepackage{threeparttable}
\usepackage{makecell}
\usepackage[font=small,skip=0pt]{caption}
\newcommand{\subparagraph}{}
\usepackage[hang]{footmisc}
\usepackage[normalem]{ulem}
\usepackage{scalerel}
\usepackage{pifont}
\usepackage{booktabs}
\usepackage{tcolorbox}
\usepackage{adjustbox}
\usepackage[caption=false, font=footnotesize]{subfig}

\theoremstyle{definition}

\theoremstyle{remark}

\makeatletter
\def\hlinew#1{\noalign{\ifnum0=`}\fi\hrule \@height #1
\futurelet\reserved@a\@xhline}
\makeatother

\definecolor{greyf}{rgb}{0.7, 0.7, 0.7}
\definecolor{greys}{rgb}{0.85, 0.85, 0.85}

\newcolumntype{a}{>{\columncolor{Gray}}c}
\newcolumntype{b}{>{\columncolor{white}}c}

\newcommand{\PreserveBackslash}[1]{\let\temp=\\#1\let\\=\temp}
\newcolumntype{C}[1]{>{\PreserveBackslash\centering}p{#1}}
\newcolumntype{R}[1]{>{\PreserveBackslash\raggedleft}p{#1}}
\newcolumntype{L}[1]{>{\PreserveBackslash\raggedright}p{#1}}

\begin{document}

% \title{DEFLECT: Towards Fast Backdoor Exposure via Latent Random Probing in Data-Free Cases}
\title{Fast and Lightweight Backdoor Detection via Head Random Probing}
% \title{HeadProber: A Unified Fast Backdoor Detection Framework for Multiple Deep Learning Tasks}
%  via Head Random Probing
% \title{One Defense for All: Fast Backdoor Detection across Diverse Deep Learning Tasks}
% \title{Universal and Efficient Backdoor Detection across Multiple Deep Learning Tasks}
\author{Yinbo Yu, \IEEEmembership{Member, IEEE}, Xueyu Yin,  Jing Fang, Chunwei Tian, \IEEEmembership{Senior Member, IEEE}, Qi Zhu, \IEEEmembership{Member, IEEE}, Jiajia Liu, \IEEEmembership{Fellow, IEEE}, and, Daoqiang Zhang, \IEEEmembership{Senior Member, IEEE}
\thanks{Y. Yu Q. Zhu, and D. Zhang are with the College of Artificial Intelligence, Nanjing University of Aeronautics and Astronautics, Nanjing, Jiangsu, 210016, China (e-mail: yinboyu@nuaa.edu.cn).}
\thanks{X. Yin and J. Liu are with the School of Cybersecurity, Northwestern Polytechnical University, Xi'an 710072, China.}
\thanks{J. Fang is with the Shenzhen Research Institute of Northwestern Polytechnical University, Shenzhen, Guangdong, 518057, China.}
\thanks{C. Tian is with the School of Computer Science and Technology, Harbin Institute of Technology, Harbin, Heilongjiang, 150001, P.R. China.}
}

\maketitle
\begin{abstract}
% Deep neural networks (DNNs) remain critically vulnerable to backdoor attacks, wherein adversaries implant hidden triggers to induce targeted misclassification upon activation.

% Image classification, object detection, and sequential decision-making
%
% backbone and head
%
% Most backdoor detection focus on recognizing trigger patterns from

Deep neural networks (DNNs) remain critically vulnerable to backdoor attacks. Existing post-training detectors often require clean or surrogate data, gradients, or iterative trigger reconstruction, leading to high computational costs and limited robustness under practical model-auditing scenarios. In this paper, we propose \textbf{HTell}, a fast and lightweight data-free backdoor detector based on head random probing. Instead of reconstructing diverse trigger patterns, HTell inspects their unified manifestation in the prediction head: backdoored models tend to exhibit abnormal response concentration on the target class under random latent probes. HTell generates architecture-aware random latent probes, feeds them directly into the model head, and detects backdoors by analyzing class-wise response statistics, without accessing real or surrogate data, model gradients, or parameter optimization.
We evaluate HTell on a large-scale benchmark containing more than 6,000 backdoored models and over 700 clean models, covering 4 datasets, 14 architectures, and 21 types of backdoor attacks. HTell achieves 99.03\% true positive rate and 2.11\% false positive rate with only 12.69 ms/model detection latency, reducing the time cost by over 30,000$\times$ compared with representative gradient-based detectors. These results demonstrate that head random probing provides an accurate, robust, and efficient solution for large-scale data-free backdoor model auditing.

\end{abstract}


\section{Introduction}
\label{sec:intro}

Deep neural networks (DNNs) have been widely deployed in safety-critical applications, yet they remain vulnerable to backdoor attacks~\cite{li2022backdoor}. A backdoored model behaves normally on benign inputs but consistently predicts an attacker-specified target label once a hidden trigger is present. This threat becomes increasingly practical with the rapid growth of third-party model sharing platforms, where users may directly download, fine-tune, or deploy pretrained models without access to their original training data or training pipeline. Since backdoor attacks usually preserve benign accuracy, efficiently inspecting released models for hidden malicious backdoors has become a critical requirement for trustworthy DNN deployment and large-scale model repository auditing.

\begin{figure}[!t]
    \centering
    \includegraphics[width=1\linewidth]{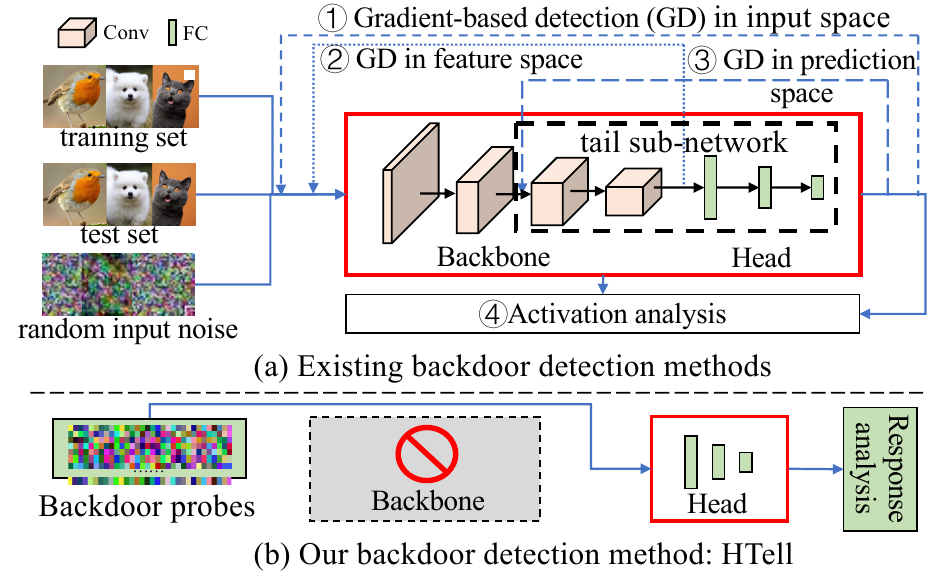}
    \caption{Method comparisons. Existing post-training methods \ding{172} (e.g., \cite{wang2019neural, wang2024mm}), \ding{173} (e.g., \cite{wang2022rethinking}), \ding{174} (e.g., \cite{fu2023freeeagle, zhang2025barbie}), and \ding{175} (e.g., \cite{chen2019detecting, tran2018spectral}) require data access or gradient backpropagation, resulting in restricted application scenarios or high detection time consumption. HTell only uses random noise to probe the model's head without gradient backpropagation, enabling fast and robust backdoor detection.}
    \label{fig:example}
    % \vspace{-5mm}
\end{figure}

Existing defenses can be broadly divided into data-level \cite{tran2018spectral, yin2025adversarial}, input-level \cite{ma2024need, zhang2025test}, and model-level \cite{wang2019neural, wang2022rethinking, wang2024mm} approaches. Data-level and input-level defenses require access to training or test samples, making them unsuitable for post-training model auditing. Recent studies \cite{fields2021trojan, fu2023freeeagle, zhou2024data, wang2024mm, zhang2025barbie} have therefore explored data-free\footnote{In these works, ``data-free'' denotes a post-training scenario where the defender has access to neither the queried model's training/test samples nor data-like surrogate samples.} model-level detection, where only the suspect model is available. Nevertheless, current detectors still face two critical limitations. \textit{First}, prevalent defenses rely on gradient-based trigger reconstruction \cite{wang2019neural, wang2022rethinking} or latent-space optimization \cite{fu2023freeeagle, zhang2025barbie}, imposing prohibitive computational overhead for large-scale models. For instance, NC \cite{wang2019neural} requires over 8 days to inspect a 250-class model \cite{ma2024need}, while the recent BARBIE \cite{zhang2025barbie} still needs more than 5 hours for a 200-class model. This inefficiency starkly contrasts with modern attacks \cite{cao2024data}, which inject backdoors into a 200-class model in just 73.3 ms. \textit{Second}, many methods exhibit low robustness \cite{wang2022rethinking, wang2020practical, fu2023freeeagle} due to sensitivity across diverse factors, including datasets, model architectures, trigger properties, poisoning rates, attack injection methodologies, and attack outputs. This asymmetry in cost and robustness poses a significant challenge to practical, large-scale model auditing.

In this paper, we address this challenge by shifting the detection focus from diverse trigger patterns to their unified manifestation in the prediction head, \textit{i.e.}, no matter what the main content is, as long as a trigger is encountered, the model will predict it as the backdoor label. While both the model's backbone and head govern its behaviors (including backdoor behavior), its prediction head critically determines backdoor manifestation \cite{fu2023freeeagle, fields2021trojan, rakin2020tbt}. Our key observation is that backdoor implantation tends to enlarge or distort the decision region of the target class in the latent space \cite{wang2022rethinking, su2024model}. Instead of characterizing decision regions with real samples through the backbone \cite{su2024model}, random latent noise (\textit{i.e.}, backdoor probes) fed directly to the head is more likely to be classified into the backdoor target class or to produce abnormally concentrated responses (see Fig. \ref{fig:example}). This head-level manifestation provides a simple and efficient signal for detecting backdoors without trigger reconstruction or access to real data.

Based on this observation, we propose \textbf{HTell}, a fast and robust backdoor detection method via head random probing. Given a suspicious model, instead of characterizing decision spaces with real samples and costly feature extraction \cite{karimi2019characterizing, somepalli2022can, su2024model}, HTell feeds isotropic random latent probes, generated following an architecture-specific distribution, directly into the head. These probes are more likely to be trapped by the backdoor decision space (see Fig. \ref{fig:tsne1}), and their output logits provide an efficient sketch of these spaces for backdoor analysis. HTell then analyzes the class-wise response statistics of the head to identify whether the model is backdoored and, if so, which class is the backdoor target. This design requires no data, gradients, or parameter optimization and makes detection more robust to heterogeneous backdoor attacks.

We conduct a large-scale evaluation to validate HTell. Our benchmark contains more than 6,000 backdoored models and over 700 clean models, covering 4 datasets (all classes), 14 network architectures (from shallow CNNs to complex vision transformers), and 21 types of backdoor attacks. These evaluated attacks cover most \textit{trigger patterns} (e.g., input- \cite{gu2017badnets, qi2023adappatch, cheng2024lotus} and feature-space \cite{nguyen2020inputaware, li2021ssba, wang2022bppattack} triggers with different poisoning rates and trigger locations, colors, textures, shapes, and sizes), \textit{attack injection methods} (e.g., training pipeline manipulation \cite{bagdasaryan2021blind}, parameter modification \cite{cao2024data}, and bit-flip \cite{rakin2020tbt, bai2022hardly}), \textit{source-target relationships} (all-to-one, all-to-all, and clean-label attack modes), and also \textit{self-supervised learning} \cite{jia2022badencoder, tao2024distribution}. Experimental results show that HTell achieves consistently strong detection performance across diverse settings. On the benchmark, HTell obtains an average true positive rate (TPR) above 99\% and a false positive rate (FPR) of 2.11\%. More importantly, HTell only takes 12.69 ms per model on average, reducing the detection latency by more than 30,000× compared with representative detectors. These results demonstrate that head random probing provides an effective accuracy-efficiency tradeoff for practical model auditing.

Our contributions are summarized as follows:

% \begin{itemize}
%   \item We analyze backdoor manifestations in the latent space of diverse backdoor models using real samples and backdoor probes. This inspires us to generate backdoor probes to expose backdoor attacks;
%   \item We propose a fast and robust backdoor detection mechanism, HTell, which only requires the model's classifier to process input backdoor probes without any information about model parameters or gradients;
%   \item We conduct a large-scale backdoor detection benchmark, including diverse datasets, model architectures, backdoor triggers, and attack strategies. Numerical results show that HTell outperforms existing methods in terms of accuracy, overhead, and speed.
% \end{itemize}

\begin{itemize}
  \item We identify a unified head-level manifestation of diverse backdoors: their target class exhibits abnormal response concentration under random latent probing;
  \item We propose HTell, a lightweight backdoor detector that inspects only the head using random probes, without real or surrogate data, gradients, or optimization;
  \item We build a large-scale benchmark covering diverse datasets, architectures, and attack strategies, and show that HTell achieves high detection accuracy with millisecond-level latency, substantially outperforming existing detectors in the accuracy-efficiency tradeoff.
\end{itemize}

% Different from \cite{su2024model}, which uses image samples to measure decision boundaries from the complex feature extractor,

% First, in the initial configuration phase, HTell coarsely characterizes the latent space using input noises to determine the architecture-specific probe distribution for inspecting models of different network architectures trained on different datasets;

% While both feature extractor and classifier govern model behavior, the classifier critically determines backdoor manifestation \cite{fu2023freeeagle, fields2021trojan, rakin2020tbt}. Model X-ray \cite{su2024model} uses image samples to measure decision boundaries for backdoor detection, which suffers from high computational overhead. HTell does not directly measure the decision boundary, but assesses the classification layer's response to latent-space noise probes (\textit{i.e.,} backdoor probes) which show heightened prediction likelihood for backdoor classes due to their expanded boundaries.

%

\section{Preliminaries and Related Work}
\label{sec:background}
\subsection{Deep Neural Networks}

In this paper, we focus on DNN models $\mathcal{M}$ for the image classification task: $\mathcal{M}: \mathbb{R}^{\mathcal{X}}\mapsto \mathbb{R}^\mathcal{Y}$, where $\mathcal{X}$ is the input space, and $\mathcal{Y}$ is the label space. A $\mathcal{M}$ has a backbone $\mathcal{B}$ for feature extraction and a head $\mathcal{H}$ for label prediction:
\begin{equation}
    \mathcal{M}(x)=\mathcal{H}(\mathcal{B}(x)),
\end{equation}
where $x\in\mathcal{X}$ is an input. We denote the latent embedding as $\mathbf{h}$ extracted from the input space $\mathcal{X}$ via $\mathcal{B}$ and $\mathcal{H}$ transforms $\mathbf{h}$ into the final prediction logits $\mathcal{M}(x)$, where $\mathbf{h}\in \mathbb{R}^D$, $D$ is the input dimensions of $\mathcal{H}$. In mainstream network architectures (e.g., GoogleNet, ResNet, ViT, VGG), $\mathcal{H}$ is their last block that consists of one or more fully-connected (FC) layers and activation functions. There are also some architectures, e.g., SqueezeNet \cite{iandola2016squeezenet}, which employs 1$\times$1 convolution as their classifier for lightweight purposes. In this paper, we follow the dimensions $D$ of $\mathbf{h}$ to generate backdoor probes $\mathbf{h}^{\text{probe}}$ and input them into $\mathcal{H}$ to expose backdoor attacks.

\subsection{Backdoor Attacks}
\label{sec:backdoorattack}

A backdoor attack aims to fine-tune parameters $\theta$ of $\mathcal{M}$ as $\overline{\theta}$ to enable behaviors including (1) keeping the benign behaviors, \textit{i.e.}, $\arg\max\mathcal{M}_{\overline{\theta}}(x)=y$, for any benign samples $x\in\mathcal{X}$; (2) classifying any poisoned sample $\hat{x}$ embedded with a trigger to a backdoor target label $t\in\mathcal{Y}$, \textit{i.e.}, $\arg\max\mathcal{M}_{\overline{\theta}}(\hat{x})=t$. The attacker typically associates poisoned samples $\hat{x}$ from any of source classes $\mathcal{Y}$ with an attack target label $t$ (\textit{i.e.,} \textit{all-to-one} attack) or those $\hat{x}$ from a specific source $s\in\mathcal{Y}$ with a corresponding target (\textit{i.e.,} \textit{all-to-all} attack). We denote $\hat{x}=\mathcal{T}(x)$, where $\mathcal{T}$ is a trigger injection function. BadNets \cite{gu2017badnets} is one of the earliest backdoor attacks, which uses a small white square $p$ as the trigger to replace a patch of $x$:
\begin{equation}\label{equ:injection}
    \hat{x}=\mathcal{T}(x)=(1-m)\odot x+m\odot p,
\end{equation}
\noindent where $m\in\{0,1\}^{C\times W\times H}$ is a binary mask and $\odot$ is element-wise multiplication. The adversary generates a poison dataset $\mathcal{D}_{poison}$ and combines it with $\mathcal{D}_{benign}$ with a poisoning rate $\lambda$ to perform backdoor implantation via model training.

Besides this above patch pattern, other patterns have also been studied, including blended \cite{chen2017blended} and perturbation \cite{nguyen2020wanet, doan2021lira, li2021ssba}. Depending on how $p$ is constructed, existing attacks can further be divided into \textit{input-space} (e.g., \cite{gu2017badnets, chen2017blended, liu2018trojannn, qi2023adappatch, cheng2024lotus}) and \textit{feature-space} backdoor (e.g., \cite{nguyen2020wanet, doan2021lira, li2021ssba, nguyen2020inputaware, wang2022bppattack, zeng2021rethinking}). The former employs ready-made images or patches as triggers. Recently, Adaptive-patch \cite{qi2023adappatch} partitions patches to generate adaptive triggers to suppress latent separability and evade detection. Similarly, LOTUS \cite{cheng2024lotus} partitions target samples and applies unique triggers to different partitions to achieve evasive and resilient attacks; The latter introduces complex trigger generation mechanisms (e.g., warping transformation \cite{nguyen2020wanet}, encoder-decoder \cite{doan2021lira, nguyen2020inputaware, li2021ssba}, GAN \cite{zeng2021rethinking}, adversarial samples \cite{wang2022bppattack}) to generate imperceptible triggers. While most backdoor attacks are \textit{sample-agnostic} attacks, \cite{li2021ssba, nguyen2020inputaware} are \textit{sample-specific} attacks.

Beyond trigger design, backdoor attacks can also differ in their implantation processes. Besides standard data poisoning \cite{gu2017badnets, lv2023data}, recent attacks inject backdoors by manipulating the training pipeline \cite{bagdasaryan2021blind, yu2023spatiotemporal} or directly modifying model parameters \cite{costales2020live, rakin2019bit, cao2024data}. For example, training-code manipulation can implant backdoors by altering the loss computation \cite{bagdasaryan2021blind}, while bit-flip attacks modify a few quantized parameters to enable runtime backdoor injection \cite{rakin2019bit, bai2022hardly}. More recently, retraining-free and data-free parameter-modification attacks can inject backdoors within less than 100 ms \cite{cao2024data}, further highlighting the need for efficient backdoor detection.

The above attacks mainly follow the dirty-label setting, where poisoned samples are assigned to an attacker-specified target label. Backdoor attacks also include clean-label variants that preserve the original source labels of poisoned samples \cite{mauro2019sig, turner2019labelconsistent, zeng2023narcissus, lv2023data}, as well as self-supervised learning (SSL) backdoors that compromise pre-trained encoders and transfer the backdoor behavior to downstream classifiers \cite{jia2022badencoder, tao2024distribution}.

\subsection{Backdoor Detection}
\label{sec:backdoordetection}

% \begin{table}
%   \caption{title}
%   \begin{tabular}
%
%   \end{tabular}
% \end{table}

% ReBack is Dataset-level defenses \cite{ma2024need}

Existing defenses against DNN backdoor attacks can be grouped into training-phase, inference-phase, and post-training methods. Training-phase defenses remove poisoned samples \cite{chen2019detecting, tran2018spectral} or modify learning procedures \cite{huang2022backdoor}, but require access to the training data and pipeline. Inference-phase defenses inspect suspicious inputs through perturbation \cite{gao2019strip}, region localization \cite{doan2020februus}, input transformation \cite{guo2023scale, liu2023detecting}, or prediction-consistency analysis \cite{ma2024need}, yet they may confuse backdoor-triggered behavior with natural model uncertainty \cite{wang2022rethinking}. This work focuses on post-training detection, where only a queried model $\mathcal{M}$ is available and the goal is to determine whether it is backdoored and identify the target label.

% Post-training detectors typically reconstruct triggers \cite{wang2019neural, wang2022rethinking, xiang2020detection, popovic2025debackdoor}, stimulate internal neurons \cite{liu2019abs}, train detection classifiers \cite{chen2019deepinspect, xu2021detecting}, or inspect statistical anomalies in outputs and activations \cite{tang2021demon, cai2022randomized}. These methods often require clean samples, training data, or gradient optimization. Recent data-free methods reduce the dependence on real data by using adversarial perturbations \cite{wang2020practical}, universal litmus patterns \cite{kolouri2020universal}, final-layer statistics \cite{fields2021trojan}, synthetic inputs \cite{fu2023freeeagle}, latent inversion \cite{zhang2025barbie}, parameter masking \cite{zhou2024data}, or maximum-margin analysis \cite{wang2024mm}. However, many of them still analyze the full model, feature extractor, or tail sub-network through per-class iterative optimization, leading to high computational overhead as model sizes and label spaces grow. The increasing efficiency of modern backdoor attacks \cite{cao2024data} further exposes this mismatch and motivates fast, lightweight, and robust data-free detection.

Existing post-training detectors mainly rely on trigger reverse engineering, neuron stimulation, detection classifiers, or statistical anomaly inspection. Representative methods reconstruct triggers in input or feature space \cite{wang2019neural, wang2022rethinking, xiang2020detection, popovic2025debackdoor}, probe internal activations to identify compromised neurons \cite{liu2019abs}, train meta-detectors for model inspection \cite{chen2019deepinspect, xu2021detecting}, or detect anomalies in outputs and channel statistics \cite{tang2021demon, cai2022randomized}. Although effective in specific settings, these methods typically require clean samples, training data, or gradient-based optimization, which limits their practicality in practical large-scale model auditing.

To relax the data requirement, recent studies have explored data-free backdoor detection. TND \cite{wang2020practical} evaluates per-class robustness using adversarial perturbations, universal litmus patterns \cite{kolouri2020universal} probe models without real data, and DQ \cite{fields2021trojan} inspects final-layer weight statistics. More recent methods, including FreeEagle \cite{fu2023freeeagle}, BARBIE \cite{zhang2025barbie}, DF-MREL \cite{zhou2024data}, and MM-BD \cite{wang2024mm}, further analyze logits, latent representations, masked parameters, or maximum-margin statistics under synthetic inputs. However, many of them still analyze the full model, feature extractor, or tail sub-network through per-class iterative optimization, leading to high computational overhead as model sizes and label spaces grow. The increasing efficiency of modern backdoor attacks \cite{cao2024data} further exposes this mismatch and motivates fast, lightweight, and robust data-free detection.

\section{Overview}
\label{sec:method}
\subsection{Threat Model}
In this paper, we have the following threat model of the attacker and defender:

% we focus on inspecting DNN models without access to any data in the black-box setting, both at the post-training phase and during inference. This data-free and black-box setting allows our method to be applied to a wider range of scenarios, including model supply-chain audit, edge-side model protection, online model monitoring, etc. Specially, we have the following threat model:

\textbf{Capability and Goal of the Attacker.} We assume that the attacker aims to inject arbitrary backdoor attacks into models via data poisoning, training process manipulation, or parameter modifications.

\textbf{Capability and Goal of the Defender}. Following \cite{zheng2022data, fu2023freeeagle, wang2024mm, zhang2025barbie}, we consider a practical data-free capability of the defender to defend against diverse backdoor attacks that the defender only has access to models without any clean or backdoor samples, or other surrogate data \cite{lv2023data}. Specifically, given a model $\mathcal{M}$, the defender only has the knowledge including its network architecture and its head $\mathcal{H}$'s input $D$ and output $\mathcal{Y}$ dimensions. To detect backdoors, the defender queries $\mathcal{H}$ as an oracle to obtain its predicted logits $z$ without access to its parameters and gradients. We further assume that the defender has no knowledge of backdoor attacks as \cite{wang2024mm, zhang2025barbie}. For $\mathcal{M}$, the defender's goal is to answer the following two questions: 1) \textbf{RQ1}: Whether $\mathcal{M}$ is benign or backdoored? 2) \textbf{RQ2}: If backdoored, which class is poisoned?
% We assume that the defender has the following knowledge about the inquired model: its network structure, the input dimensionality, the number of classes, and the output logits of its classification layer $\mathcal{H}$ given an arbitrary input query. This assumption requires the model owner to provide an interface with only access to the classification layer, so that the input request can be processed without the need to run a feature extractor.

% This access can be provided by model-sharing or model-selling platforms like Model Zoo and AWS Marketplace, as well as model infrastructure managers for model inspections. For backdoor detections, the defender does not require high computational resources (e.g., GPU), access to any training and test datasets, or to collect any images. This setting requires the backdoor detection to be conducted in a low-overhead but efficient way. In addition, we assume that the defender is also the platform manager who can accumulate a set of clean model samples of typical networks trained under different datasets. This is easy to accumulate in advance.

\subsection{Key Intuition and Idea}
\label{sec:key}

% the backdoor behavior must eventually be reflected in the decision geometry of $\mathcal{H}$.

Our key idea is to detect the unified head-level manifestation of backdoors rather than reconstructing diverse trigger patterns in the input or feature space. For a backdoored model $\mathcal{M}$, a triggered sample $\hat{x}$ is expected to be classified into the attacker-specified target label $t$: $\arg\max \mathcal{H}(\mathbf{h}_{\hat{x}}) = t$, where $\mathbf{h}^{\hat{x}}$ is the latent feature of $\hat{x}$. Since the head $\mathcal{H}$ directly determines the final logits, it provides a compact and efficient place to inspect the backdoor behavior. For a linear head, the logit of class $i$ can be written as:
\begin{equation}
    z_i = \mathbf{w}_i^\top \mathbf{h} + b_i, \quad i\in \mathcal{Y} .
\end{equation}

To ensure a successful attack, the target logit should dominate other logits for poisoned features:
\begin{equation}\label{equ:margin}
    (\mathbf{w}_t^\top \mathbf{h}^{\hat{x}} + b_t) -
    (\mathbf{w}_j^\top \mathbf{h}^{\hat{x}} + b_j) > \Delta, \quad \forall j \neq t ,
\end{equation}
where $\Delta>0$ is a classification margin. In the latent space, $\mathbf{h}^{\hat{x}}=\mathbf{h}^x+\delta$, where $\delta$ is the backdoor perturbation introduced by triggers \cite{wang2022rethinking}. So, Equ. (\ref{equ:margin}) can be transformed as:
\begin{equation}\label{equ:backdoor}
    \underbrace{(\mathbf{w}_t^\top - \mathbf{w}_j^\top)\mathbf{h}^x}_{\text{Clean classification}} + \underbrace{(\mathbf{w}_t^\top - \mathbf{w}_j^\top) \delta + (b_t - b_j)}_{\text{Backdoor perturbation}} \geq\Delta, \quad \forall j \neq t.
\end{equation}
\noindent In the second item, $\mathbf{w}_t^\top\delta=\|\mathbf{w}_t\|_2\cdot \|\delta\|_2\cdot\cos{\varphi^t}$, where $\varphi^t$ is the angle between $\mathbf{w}_t$ and $\delta$. As studied in \cite{wang2022rethinking, zhang2024exploring}, $\delta$ tend to be orthogonal to clean features, leading to an initial large $\varphi^t$ or even $\cos{\varphi^t}\approx0$. Therefore, to achieve Equ. (\ref{equ:backdoor}), backdoor implantation may update $\mathcal{H}$ through different parameter changes, such as increasing the target weight norm $\|\mathbf{w}_t\|_2$, adjusting partial target-related dimensions, or modifying the target bias $b_t$. These changes vary across datasets, architectures, triggers, and attack strategies, which explains why detectors relying on a single parameter statistic (like \cite{fields2021trojan}) can be unstable.

\begin{figure*}[t!]
\centering

\begin{minipage}[c]{1\textwidth}
    \centering
    \includegraphics[width=0.55\textwidth]{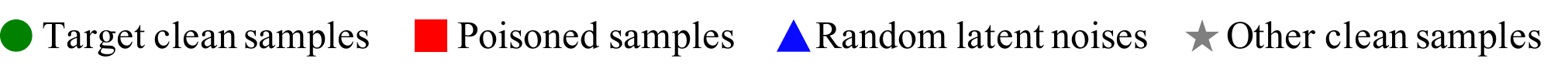}
  \end{minipage}
  \vspace{-6mm}

\begin{minipage}[c]{1\textwidth}
    \centering
    \subfloat[Badnet \cite{gu2017badnets}]{\includegraphics[width=0.195\textwidth]{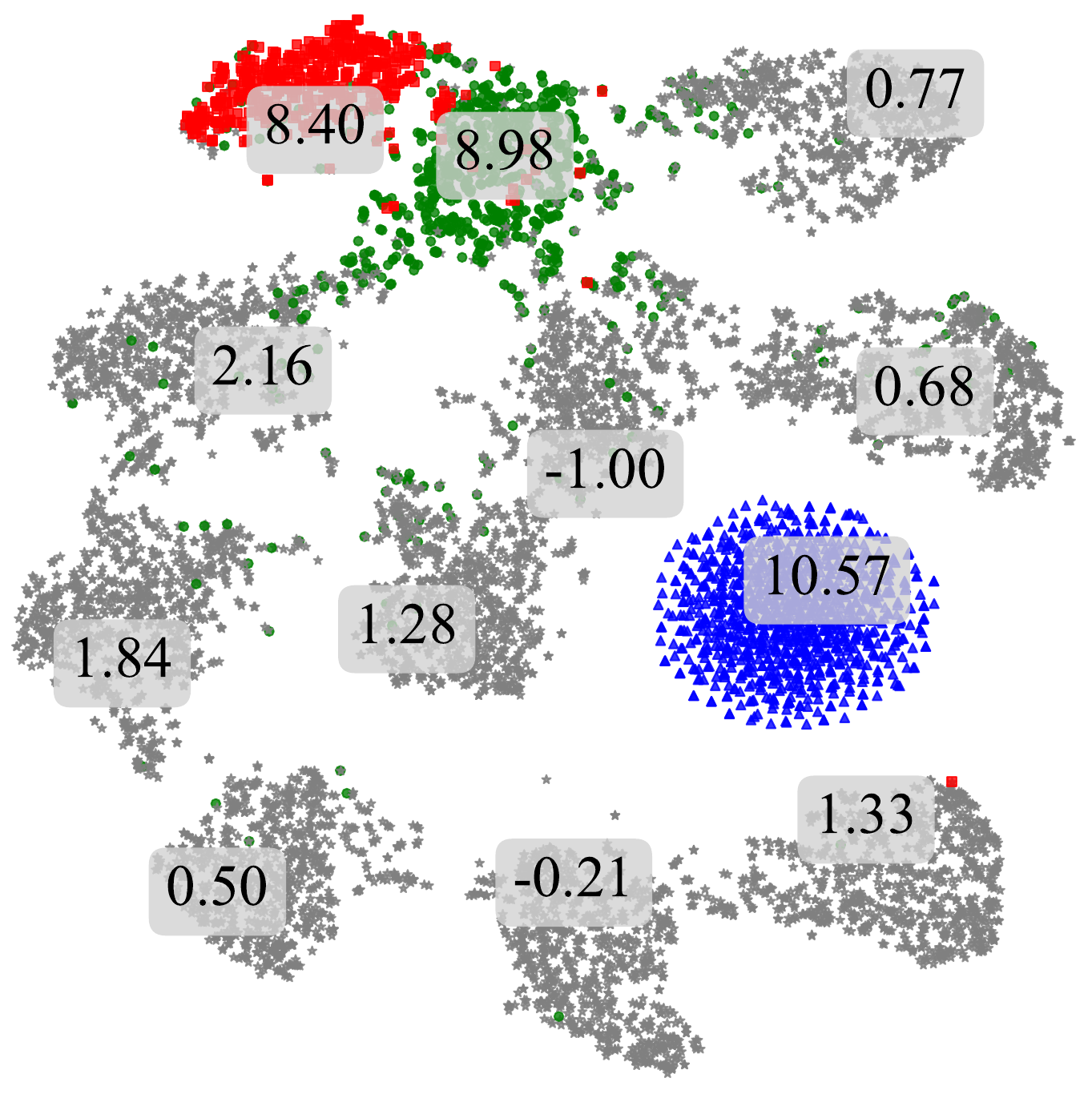}}
    \hfill
    \subfloat[Blended \cite{chen2017blended}]{\includegraphics[width=0.195\textwidth]{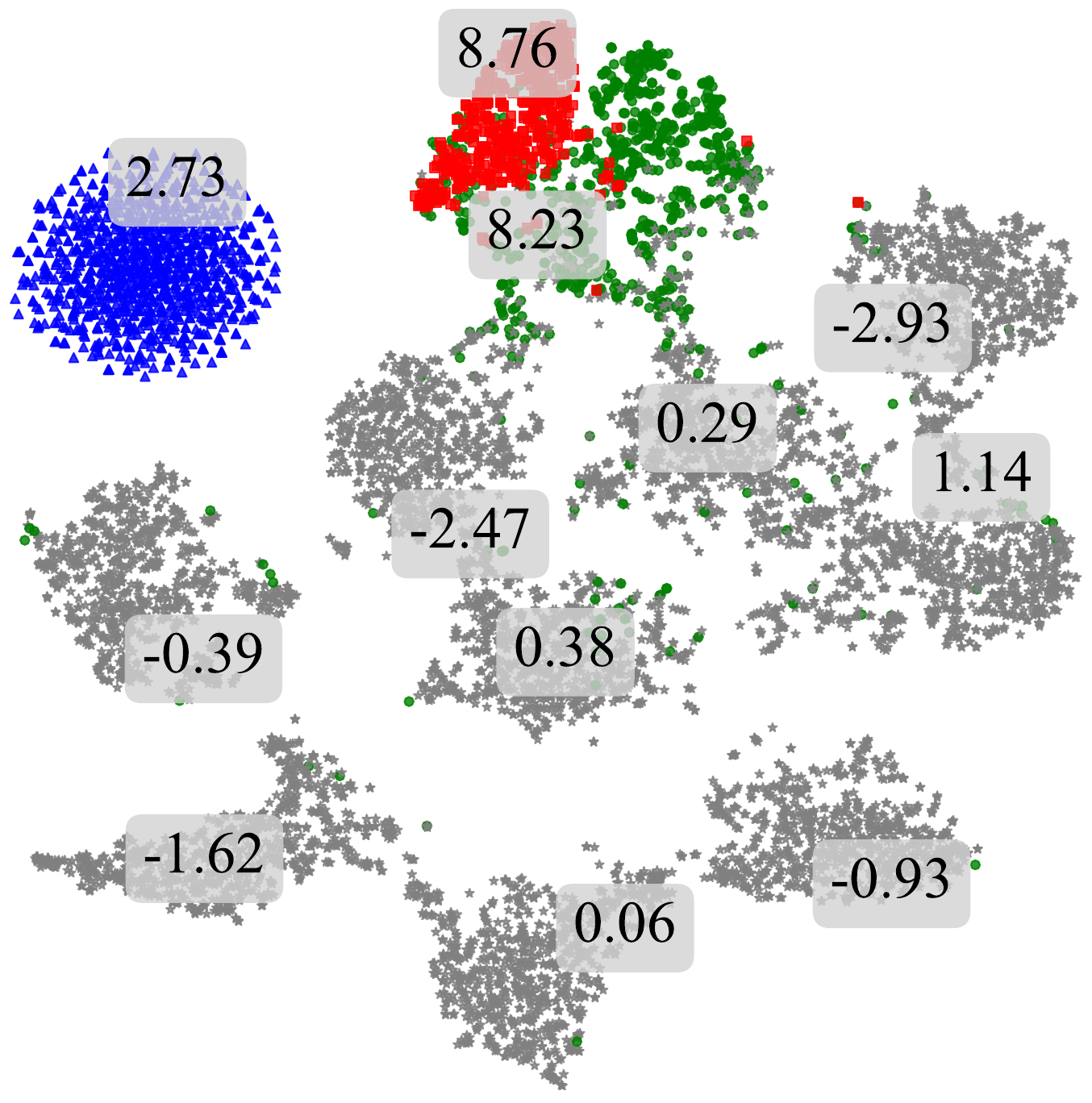}}
    \hfill
    \subfloat[BppAttack \cite{wang2022bppattack}]{\includegraphics[width=0.195\textwidth]{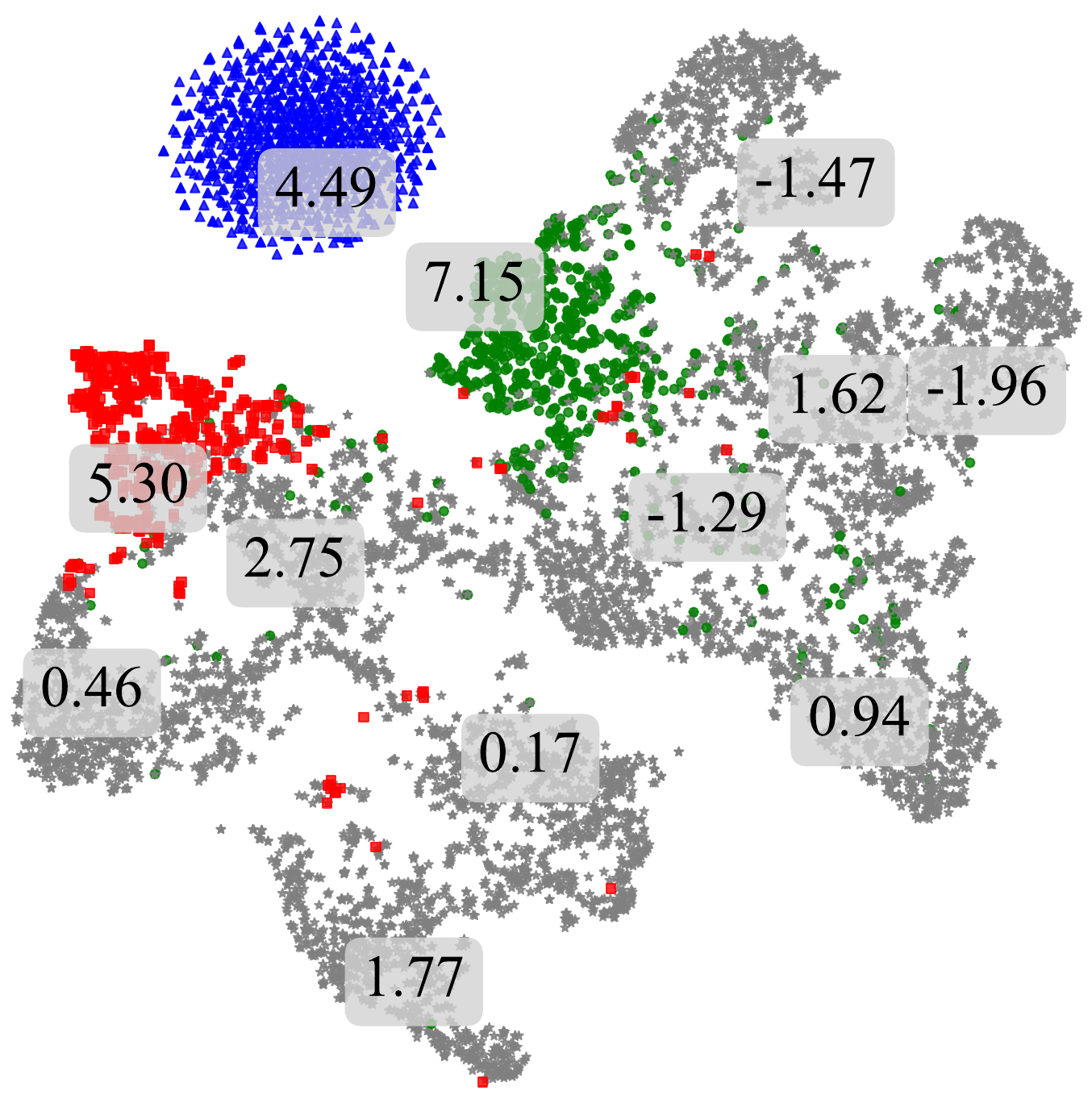}}
    \hfill
    \subfloat[Inputaware \cite{nguyen2020inputaware}]{\includegraphics[width=0.195\textwidth]{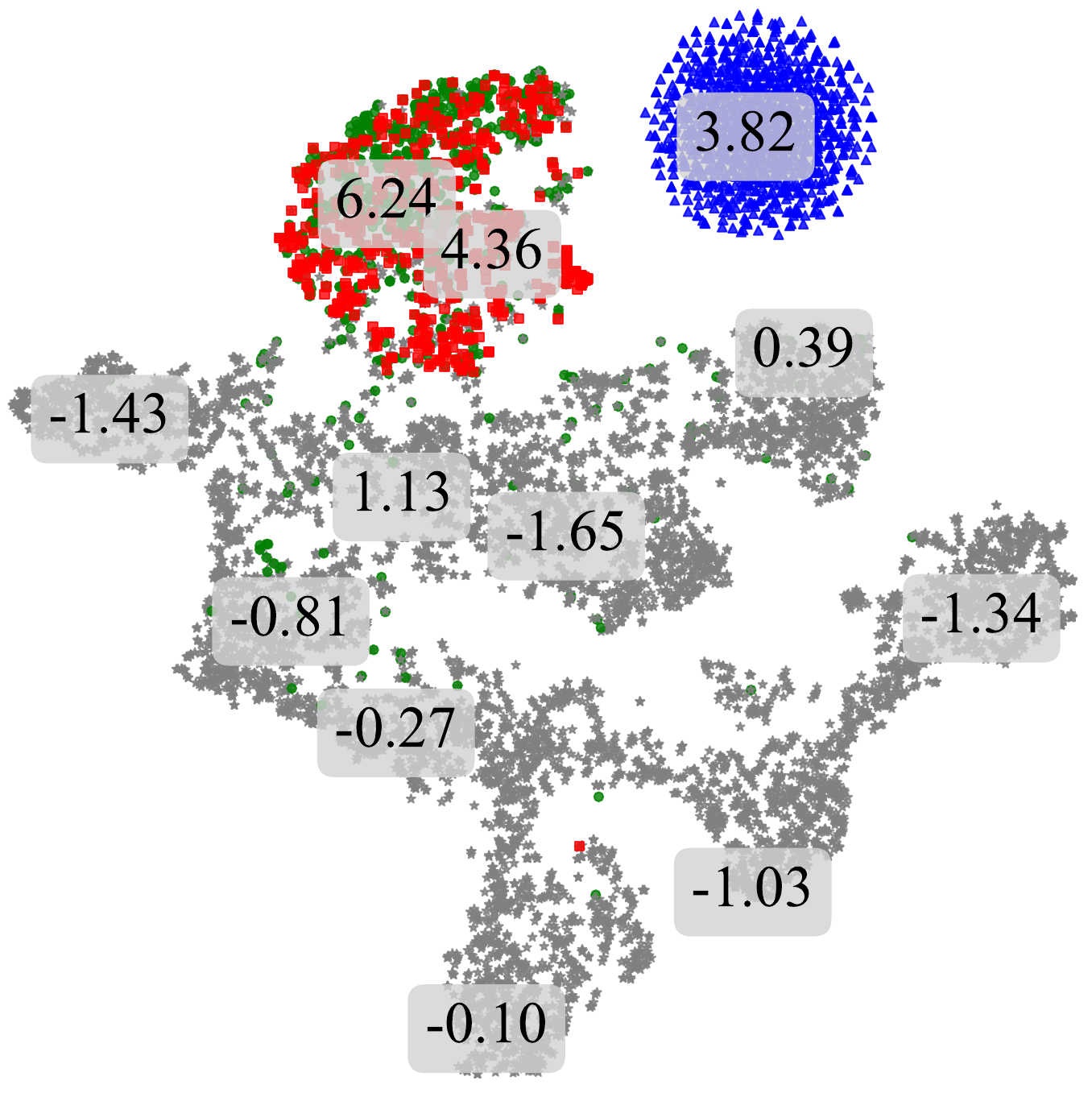}}
    \hfill
    \subfloat[ISSBA \cite{li2021ssba}]{\includegraphics[width=0.195\textwidth]{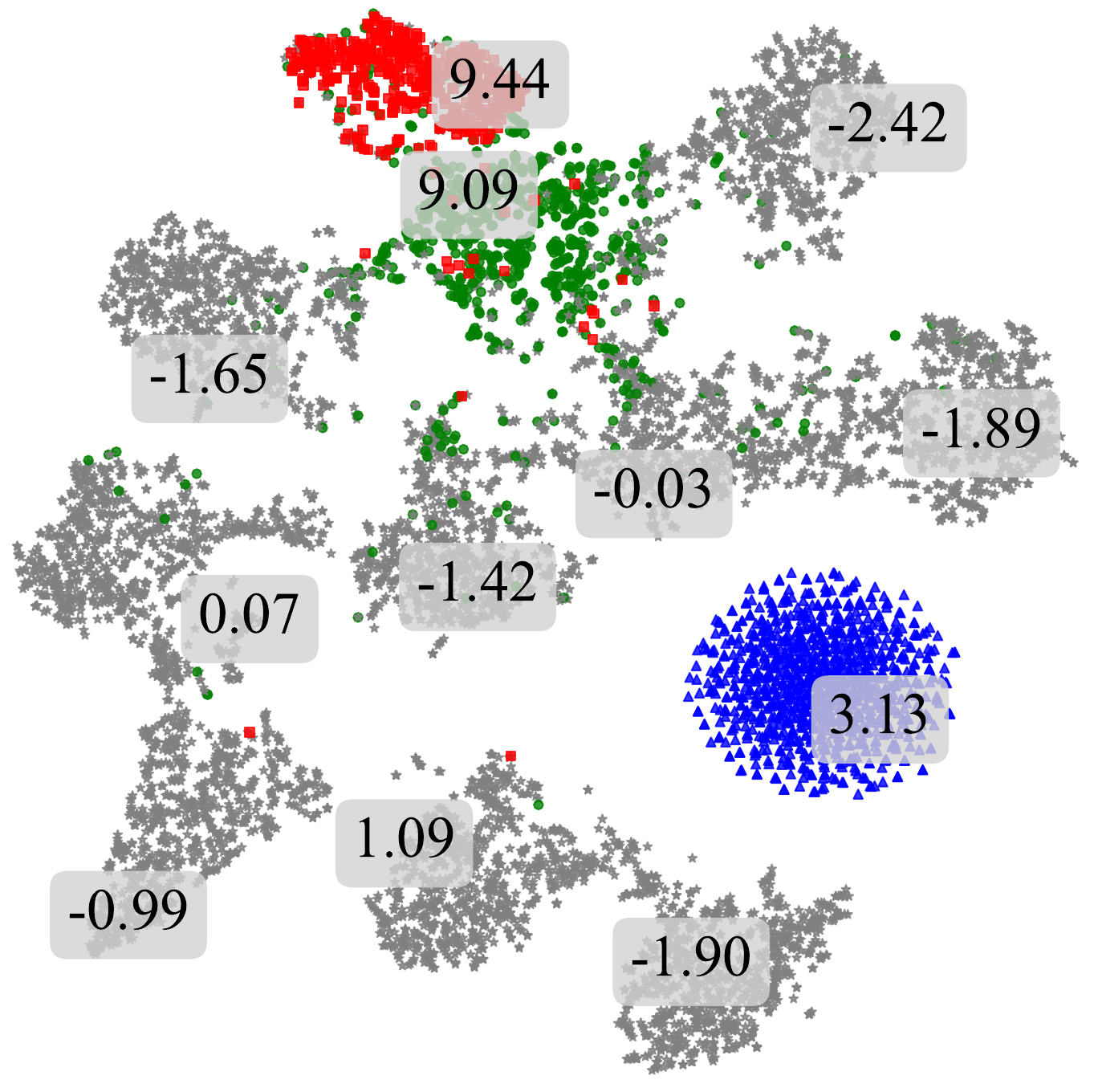}}
  \end{minipage}
  % \vspace{-2mm}

% \begin{minipage}[c]{1\textwidth}
%     \centering
%     \subfloat[Low Frequency \cite{zeng2021rethinking}]{\includegraphics[width=0.22\textwidth]{figures/cifar10_googlenet_lf_t4_tsne.pdf}}
%     \hspace{0.02\textwidth}
%
%     \hspace{0.02\textwidth}
%     \subfloat[Blind \cite{bagdasaryan2021blind}]{\includegraphics[width=0.22\textwidth]{figures/cifar10_googlenet_blind_t4_tsne_1.pdf}}
%     % \hspace{0.005\textwidth}
%   \end{minipage}

    % \subfloat[Trojannn \cite{liu2018trojannn}]{\includegraphics[width=0.245\textwidth]{figures/cifar10_googlenet_trojannn_t4_tsne.pdf}}
	\caption{T-SNE visualization of clean and poisoned latent features extracted by backdoored models and random backdoor probes. All backdoored models are trained on the CIFAR10 dataset using GoogleNet, with class 4 designated as the backdoor target label. The latent features of \textcolor{red}{poison}, \textcolor{green}{target clean}, and \textcolor{gray}{other clean} samples are output by $\mathcal{B}$. \textcolor{blue}{Backdoor probes} are generated by sampling from a uniform distribution $\text{U}[0,1]$. The numerical value on each cluster represents the logit $\mathbf{w}_t^\top\mathbf{h}$ of a random sample $\mathbf{h}$ with the target weight $\mathbf{w}_t$.}
	\label{fig:tsne1}
 \vspace*{-3mm}
\end{figure*}

Instead of identifying a specific parameter anomaly, we focus on the functional consequence of backdoor implantation: the target class tends to obtain an enlarged or distorted decision region in the latent space \cite{wang2022rethinking, su2024model}. Therefore, isotropic random latent probes are more likely to fall into or strongly activate the target region. Given a set of random probes $\mathbf{h}^{probe}$, a backdoored head is expected to produce abnormal response concentration on the target class, while a benign head should yield more balanced responses across classes. This observation motivates HTell to use random latent probes as lightweight ``backdoor probes'' for head-level response analysis.

% We empirically validate this intuition using backdoored GoogleNet models trained on CIFAR10. As illustrated in Fig.~\ref{fig:tsne1}, random probes (following a uniform distribution $\text{U}[0,1]$) produce target-class logits that are much larger than those of non-target clean samples and are close to those of target clean or poisoned samples across 5 different attacks, including both latent-separable and latent-inseparable triggers. This observation motivates HTell to perform fast and data-free backdoor detection by probing only the prediction head.

We empirically validate this intuition using backdoored models generated by 8 representative attacks, including BadNets \cite{gu2017badnets}, Blended \cite{chen2017blended}, BppAttack \cite{wang2022bppattack}, Inputaware \cite{nguyen2020inputaware}, and ISSBA \cite{li2021ssba}. We collect latent features of clean and poisoned samples, together with random probes $\mathbf{h}^{\text{probe}}\sim \mathrm{U}[0,1]$, and visualize them using t-SNE. As shown in Fig.~\ref{fig:tsne1}, across diverse trigger patterns, implantation strategies, and latent separability properties, the target-class logits of random probes, $\mathbf{w}_{t}^{\top}\mathbf{h}^{\text{probe}}$, are consistently much larger than those of non-target clean samples and close to those of target clean or poisoned samples. This indicates that random probes tend to activate the enlarged backdoor target decision region, motivating HTell to perform fast and robust detection via head-level probing.

\section{Detailed Methodology}
\subsection{Backdoor Probe Generation in Latent Space}
% \begin{equation}
%     p(y \mid \mathbf{h}) = \text{softmax}(z_y) =\frac{e^{\mathbf{w}_y^{\top} \mathbf{h} + b_y}}{\sum_{i=1}^{\mathcal{Y}} e^{\mathbf{w}_i^{\top} \mathbf{h} + b_i}}
% \end{equation}
Given a latent feature $\mathbf{h} \in \mathbb{R}^D$, the prediction $p$ for a class $y$ is $p(y \mid \mathbf{h}) = \text{softmax}(z_y)$.
% As discussed in Sec. \ref{sec:key}, a backdoor probe $\mathbf{h}^{\text{probe}}$ is more likely to have a prediction $p(t\mid\mathbf{h}^{\text{probe}})$ of backdoor target $t$ higher than the ones of other classes.
In this paper, we use a \textit{backdoor probe} $\mathbf{h}^{\text{probe}}$ to denote a random noise generated in the latent space according to a certain distribution, which will be input in the head $\mathcal{H}$. To ensure the success rate of backdoor probes in activating backdoor manifestations, we need to know the latent space to determine the distribution (e.g., normal or uniform) for backdoor probe generations. In general, datasets and network architectures jointly shape the distribution range of latent features and their morphology. However, in image processing tasks, images are normalized before being input into models. In this case, the effect of network architecture is more significant and direct than the dataset differences. In modern architecture, in addition to linear and convolution, there are also many other components, such as activation, normalization, pooling, \textit{etc}. These components directly and strongly shape the range and statistical properties (e.g., means, variance) of hidden features $\mathbf{h}$. Directly analyzing the network architecture to obtain the latent space is often complex.

To determine the probe distribution without real data, HTell first queries the backbone $\mathcal{B}$ with a small batch of random input noises $x^{\text{noise}}\sim\mathcal{U}[0,1]^{b\times C\times W\times H}$ and uses the resulting latent activations to estimate a coarse activation range $\mathcal{S}^{\text{noise}}$. We observe that the latent spaces of modern architectures mainly follow two patterns: non-negative activations induced by ReLU-like functions, and signed or approximately normalized activations induced by normalization layers. Accordingly, HTell adopts two probe distributions as follows.

For models whose latent activations are non-negative, HTell uses uniform probes $\mathcal{U}[0,1]$. This choice matches the activation range of common ReLU-based heads and provides sufficient coverage for architectures such as VGG, GoogleNet, and ResNet. Although the upper bound can be estimated from $\mathcal{S}^{\text{noise}}$, we empirically find that $\mathcal{U}[0,1]$ is sufficient across datasets, architectures, and attacks.

% A non-negative range is typically constrained by activation functions. For example, ReLU and its variants apply a hard lower bound (0 or close to 0), resulting in the values of its output $\mathbf{h}^{x^{\text{noise}}}$ that are non-negative and have no upper bound (\textit{i.e.}, $[0,+\infty)$). This output is skewed and sparse: a large number of values are close to 0 (corresponding to suppressed negative inputs), and the rest of the values are distributed over a wide positive interval. Hence, for this non-negative latent space, we use the uniform distribution $\text{U}[\alpha,\beta]$ to generate backdoor probes. We set $\alpha=0$ due to the constraint of ReLU, and $\beta$ can be set according to $\mathcal{S}^{\text{noise}}$. But, from our experimental results, $\beta=1$ can provide a sufficient detection performance across backdoor attacks for most network architectures on different datasets, e.g., VGG, GoogleNet, ResNet. Hence, \textit{for a $\mathcal{M}$ whose latent feature has a non-negative value range, we employ $\text{U}[0,1]$ to generate backdoor probes (\textit{i.e.}, uniform backdoor probes)}.

For models whose latent activations have a signed range, HTell uses Gaussian probes $\mathcal{N}(0,\sigma^2)$. The mean is set to zero because the noise-activated latent features are approximately centered due to normalized layers (e.g., batch, layer, group normalization). $\sigma$ is selected according to the observed range of $\mathcal{S}^{\text{noise}}$. We determine $\sigma$ according to the three sigma granularities: for $x\sim\mathcal{N}(\mu,\sigma^2)$, $P\{|x-\mu|\leq3\sigma\}\approx0.9974$, where $\mu=0$. Hence, $\sigma\geq \frac{1}{3}|\mathcal{S}^{\text{noise}}|^{\max}$. This Gaussian probing strategy better matches normalized latent spaces and can cover both positive and negative activation directions.

There are also architecture-specific exceptions. For example, SqueezeNet \cite{iandola2016squeezenet} uses a convolutional prediction head followed by ReLU and global average pooling (GAP), whose response statistics differ from standard fully-connected heads. The GAP operation averages spatial response maps and can smooth out the bounded variations induced by uniform probes, while ReLU preserves only positive pre-activation responses. Gaussian probes with a larger variance are more likely to generate strong positive local responses that survive ReLU and remain visible after pooling. Thus, although SqueezeNet has non-negative latent activations, HTell uses Gaussian probes $\mathcal{N}(0, \sigma^2)$ to better expose target-class response concentration.

% Note that the distribution determination method is designed for the most advanced network architectures whose use FC layers as their head. However, there are also networks (like SqueezeNet \cite{iandola2016squeezenet}) that employ convolution to construct their head. Specifically, SqueezeNet employs a 1$\times$1 convolution, ReLU, and a global average pool (GAP) as its head. It results in a different expectation and variance of $z_i$ with FC layers due to the truncation effect of ReLU \cite{hayou2019impact} and the pooling effect of GAP. Using the uniform distribution to activate backdoors requires abnormal spikes or $\mathbf{w}_t$ to be extremely large, which is not practical. In contrast, using a normal distribution with a higher variance $\sigma^2$ can produce positive tail spikes to compensate for the truncation effect of ReLU, thereby more likely activating backdoors. Therefore, although the latent value range of SqueezeNet is non-negative, we use a $\mathcal{N}(0, \sigma^2)$ to generate backdoor probes.
% In Appendix Section \ref{apxs:latentdistribution}, we give a more detailed analysis from perspective of network architectures to answer why these backdoor probes from different distributions are valid for backdoor detection.

\subsection{Backdoor Detection via Head Probing}
\label{sec:detectionatc}
Given a set of backdoor probes $\mathbf{H}^{\text{probe}}$ with the size of $N$, the queried head $\mathcal{H}$ outputs their predicted logits $\mathbf{z}(\cdot|\mathbf{H}^{\text{probe}})=[z_{n,i}], n\in N, i\in \mathcal{Y}$, probabilities $\mathbf{P}(\cdot|\mathbf{H}^{\text{probe}})=[p_{n,i}]$ for all classes $\mathcal{Y}$, and corresponding labels $\hat{Y}=[\hat{y}_n]=[\arg\max_ip_{n,i}]$. We consider the following 2 statistical indicators for backdoor detections:

\begin{itemize}
    % \item \textbf{Class prediction proportion}:
    % \begin{equation}
    % \mathbf{r}^{\text{ratio}}=[r_i]_\mathcal{\mathcal{Y}}=[\frac{1}{N}\sum_{n=1}^N\mathbb{I}(\hat{y}_n=i)],
    % \end{equation}
    % It can intuitively reflect the model's bias towards backdoor classes. Since random backdoor probes are isotropic or uniform, they are more likely to activate the backdoor class;
    % \item \textbf{Class maximum confidence}:
    % \begin{equation}
    % \mathbf{r}^{\text{max}}=[z_i^{\max}]_\mathcal{\mathcal{Y}}=[\max_{n=1}^Nz_{n,i}]
    % \end{equation}
    % This indicator evaluates the largest predicted logits of backdoor probes on each class. Since backdoor implantation may increase the backdoor target class's weights or bias \cite{fields2021trojan}, the target may result in larger logits than other clean classes. Hence, backdoor probes may achieve the largest predicted logits on the  target class;
    % reflecting the upper strength of the activation of the backdoor class
    \item \textbf{Class average confidence}:
    \begin{equation}
        \mathbf{r}^{\text{mean}}=[\bar{p}_i]_\mathcal{\mathcal{Y}}=[\frac{1}{N}\sum_{n=1}^Np_{n,i}]
    \end{equation}
    This indicator reflects the average activation intensity of backdoor probes on each class. $\mathbf{H}^{\text{probe}}$ may achieve the maximum average prediction probability of the backdoor target compared to other clean classes;
    \item \textbf{Class probability norm}:
    \begin{equation}
        \mathbf{r}^{\text{l2}}=[\|\mathbf{p}_{:,i}\|_2]_\mathcal{\mathcal{Y}}=[\sqrt{\sum_{n=1}^Np_{n,i}^2}]
    \end{equation}
    It quantifies distributional concentration and outlier intensity in prediction outputs. Considering the diversity of weight updates induced by backdoor implantation, backdoor models may be only sensitive to latent features in certain directions. Due to isotropy, a subset of normal backdoor probes will strongly activate the backdoor.
\end{itemize}

To detect backdoors, we choose one indicator $\mathbf{r}$ with a threshold $\tau$ according to the network architecture. Given a queried model $\mathcal{M}$, HTell generates and sends a set of random probes $\mathbf{H}^{\text{probe}}$ to its head $\mathcal{H}$. Then, it calculates $\mathbf{r}(\mathbf{P}(\cdot|\mathbf{H}^{\text{probe}}))$. We claim that if $\max_{y}^{\mathcal{Y}}\mathbf{r}>\tau$, $\mathcal{M}$ is backdoored to answer \textbf{RQ1} and the class associated with $\mathbf{r}_{\max}$ is inferred as the backdoor target class to answer \textbf{RQ2}.

\section{Evaluation}
\label{sec:eva}

% To evaluate the effectiveness and efficiency of HTell, we design the following research questions:
% \begin{itemize}
%     \item \textbf{RQ1}: How are backdoor probes similar to latent features of real clean and poison samples?
%     \item \textbf{RQ2}: How effective are detection indicators on the classification probabilities of backdoor probes?
%     \item \textbf{RQ3}: How efficient and accurate is HTell for detecting different backdoor attacks compared with other SOTA approaches?
%     \item \textbf{RQ4}: How general is HTell in different settings of datasets and network architectures?
%     \item \textbf{RQ5}: How can the attacker bypass HTell?
% \end{itemize}
%
% To answer RQ1, we aim to analyze the effectiveness of using random noise for backdoor detection. This question helps explain the essential basis of our proposed data-free detection method with practical phenomena. To answer RQ2, we analyze the impact of model architecture, dataset, and backdoor attacks on the accuracy of these detection indicators, so as to provide a more robust configuration for backdoor detection. We set RQ3 and RQ4 to guide our experiments to provide a comprehensive and large-scale performance analysis of HTell compared with existing methods. To answer RQ5, we investigate possible adaptive attacks against HTell.

\subsection{Experiment Setup}

\subsubsection{Benchmark Overview}

To comprehensively evaluate HTell under diverse attack settings, we construct a large-scale benchmark covering different datasets, architectures, trigger patterns, attack injection strategies, and source-target relationships. Compared with existing evaluations that typically consider only a few dataset-architecture pairs \cite{wang2024mm, fu2023freeeagle, mo2024robust, wang2022rethinking, wang2019neural, zhang2025barbie}, our benchmark contains more than 6,000 backdoored models and over 700 clean models.

\noindent\textbf{Datasets and Architectures.}
We evaluate HTell on four widely used datasets with different label-space scales and image resolutions, including MNIST, CIFAR10, GTSRB, and TinyImageNet. To examine architecture-level stability, we use dataset-specific model sets. For MNIST, we use CNN2~\cite{gu2017badnets} and LeNet5. For CIFAR10 and GTSRB, we use CNN6, EfficientNet-B3 (ENb3), GoogleNet (GNet), ResNet18 (R18), PreactResNet18 (PR18), VGG16-bn (V16), and SqueezeNet (SNet). For TinyImageNet, we use PreactResNet18, MobileNet-v3-large (MNet), VGG19-bn (V19), and ViT-B/16 (VIT). In addition, to evaluate the influence of model scale, we include ResNet34 and ResNet50 together with ResNet18 in the network-size and poisoning-rate stress tests. Overall, the benchmark covers 14 architectures spanning shallow CNNs, residual networks, efficient CNNs, convolutional heads, and vision transformers.
%
%
% \noindent\textbf{Datasets and Architectures.}
% We evaluate HTell on MNIST, CIFAR10, GTSRB, and TinyImageNet. The architecture set is dataset-specific: MNIST uses CNN2~\cite{gu2017badnets} and LeNet5; CIFAR10 and GTSRB use CNN6, EfficientNet-B3, GoogleNet, ResNet18, PreactResNet18, VGG16-bn, and SqueezeNet; TinyImageNet uses PreactResNet18, MobileNet-v3-large, VGG19-bn, and ViT-B/16. We additionally include ResNet34 and ResNet50 in the network-size and poisoning-rate stress tests. Overall, the benchmark covers 14 architectures with different depths, operators, and head structures.

% For VGG16-bn on CIFAR10 and GTSRB, we set the hidden dimension of its classifier to 512.

\begin{figure}
    \centering
    \includegraphics[width=0.8\linewidth]{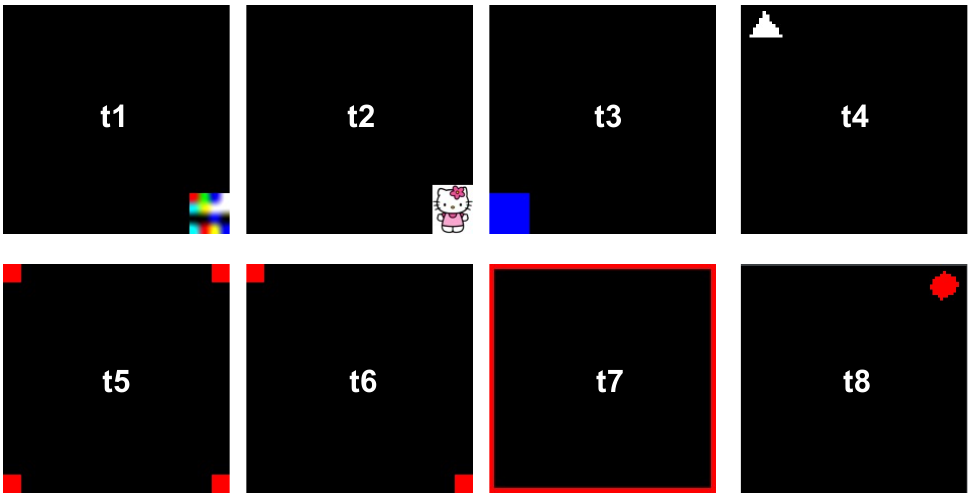}
    \caption{Badnets with different patch triggers.}
    \label{fig:trigger}
    \vspace{-2mm}
\end{figure}

\noindent\textbf{Backdoor Attacks}. We consider 21 representative attacks spanning diverse trigger patterns and implantation strategies, including patch-based attacks (BadNets, TrojanNN, Adaptive-patch, LOTUS), blended and perturbation-based attacks (Blended, WaNet, BppAttack, Lira, ISSBA, InputAware), training-pipeline manipulation (Blind), clean-label attacks (SIG, LC, Narcissus, DataFree), SSL backdoors (BadEncoder, DRUPE), parameter-modification attacks (DFBA), and bit-flip attacks (TBT, HPT). In patch-based attacks, besides white patches, we also introduce 8 new patches with various patching locations, colors, and textures (see Fig. \ref{fig:trigger}).

To reduce the effect of class imbalance on decision-space analysis, we implant attacks into different target labels across datasets. The main benchmark contains 4,850 backdoored models and 426 clean models for held-out evaluation, while additional stress tests cover all-to-all attacks (700 backdoor models), adaptive patch attacks, poisoning-rate variations, network-size variations, SSL backdoors, and parameter-modification attacks (totally 1,316 backdoor and 164 clean models). All evaluated backdoored models maintain high benign accuracy and attack success rate, ensuring valid attack implantation across different datasets and architectures.

% Besides, we trained 5 all-to-all backdoor models of each above combination of attack, network architecture, and dataset (CIFAR10 and GTSRB) and generated a total of 700 all-to-all backdoor models as an all-to-all test benchmark. For other triggers and backdoor attacks, we further trained several specific test benchmarks, including 1316 backdoor models and 164 clean models of other network architectures. In summary, we constructed a large-scale backdoor detection benchmark with over 6,000 backdoor models and over 700 corresponding clean models to cover the impact of datasets, network architectures, trigger patterns, attack injections, and poisoning rates.

\subsubsection{Evaluation Metrics}

We evaluate backdoor models using attack success rate (ASR) and benign accuracy (BA). Following prior work \cite{fu2023freeeagle, mo2024robust, wang2024mm}, we report the true positive rate (TPR), false positive rate (FPR), and average precision(AP) of backdoor detection. We additionally measure the average detection time per model to evaluate efficiency. All models are trained and detected on a server equipped with an Intel i9-12900K CPU and two NVIDIA RTX 3090 GPUs.

\subsubsection{Implementation Details}

We employ BackdoorBench \cite{wu2022backdoorbench} with default settings to train models. Unless otherwise specified, models are trained with SGD for 50 epochs using batch size 128, learning rate 1e$^{-3}$, and poisoning rate 0.1. Similar to existing detectors \cite{fields2021trojan, fu2023freeeagle, zhang2025barbie}, we train a small configuration set (85 clean and 485 backdoored models with random targets, less than \cite{fu2023freeeagle, zhang2025barbie}) to calibrate the probe distribution, response indicator, and threshold for each architecture. We estimate the coarse latent range by forwarding random input noises ($x^{\text{noise}}\sim\mathrm{U}[0,1]^{b\times C\times W\times H}$ with $b=32$) through the backbone. Architectures with non-negative latent ranges use uniform probes, while those with signed latent ranges use Gaussian probes. We use $\mathbf{r}^{\text{mean}}$ for most architectures and $\mathbf{r}^{\text{l2}}$ for PreactResNet18. The detection threshold is selected to maximize TPR while keeping FPR below 5\%.

% The configuration set is only used once for architecture-level calibration and does not contain any training or test samples of the queried model. During inspection, HTell only requires forward queries to the prediction head.
% To estimate the latent feature range, we query each model in the set using random input noises $x^{\text{noise}}\sim\mathrm{U}[0,1]^{b\times C\times W\times H}$ with $b=32$. Based on the resulting latent distributions, HTell employs either uniform or Gaussian probes depending on the architecture type. We use $\mathbf{r}^{\text{mean}}$ for most architectures and $\mathbf{r}^{\text{l2}}$ for PreactResNet18. Detection thresholds are selected to maximize TPR while keeping FPR below 5\%.

% Detailed probe distributions, thresholds, and sensitivity analyses are provided in the supplementary material.

\subsubsection{Baselines}

We compare HTell with seven representative post-training backdoor detectors, including NC \cite{wang2019neural}, FeatureRE \cite{wang2022rethinking}, DF-TND \cite{wang2020practical}, DQ \cite{fields2021trojan}, FreeEagle \cite{fu2023freeeagle}, MM-BD \cite{wang2024mm}, and BARBIE \cite{zhang2025barbie}. Among them, NC and FeatureRE require access to data, while most other baselines rely on gradient-based optimization. To ensure fairness, we use the official implementations and re-tune their detection thresholds for each dataset-architecture pair. Following \cite{wang2024mm}, we reduce the search complexity of NC, FeatureRE, and DF-TND on TinyImageNet and GTSRB by estimating only the putative target class. Since BARBIE only determines whether a model is backdoored, its reported TPR corresponds to backdoor-model detection accuracy rather than exact target-label identification.

% Due to the time constraint, we extract a small detection benchmark from our full benchmark to evaluate these baselines. Specifically, we include all models of MNIST; For CIFAR10, GTSRB, and TinyImageNet, we randomly select 11, 13, and 11 clean models per architecture and 3, 12, 10 backdoor models per attack-architecture pair, respectively. There are 232 clean models and 1350 backdoor models in this benchmark. Furthermore, with our configuration set, we further involve 65 clean models and 970 backdoor models from remained models to configure these baselines.

Due to the high computational cost of several baselines, we construct a small benchmark for baseline evaluation. Specifically, we include all MNIST models. For CIFAR10, GTSRB, and TinyImageNet, we randomly select 11, 13, and 11 clean models per architecture and 3, 12, and 10 backdoored models per attack-architecture pair, respectively, from the full benchmark. This results in 232 clean models and 1,350 backdoored models for baseline testing. For baseline calibration, we randomly select 65 clean models and 970 backdoored models from remained models in the full benchmark and combine them with the configuration set to tune the detection thresholds of baselines.

\begin{table*}[t!]
\centering
\footnotesize
\caption{Performance comparisons on different datasets (backdoor/benign models). TPR/FPR are in percentages (\%). The time is the average detection time per model.}
\label{tab:comparsion}
\begin{threeparttable}
\begin{tabular}{c|cc|cc|cc|cc|cc}
\toprule
\multirow{2}{*}{Method} & \multicolumn{2}{c|}{MNIST (20/20)} & \multicolumn{2}{c|}{CIFAR10 (210/77)} & \multicolumn{2}{c|}{GTSRB (840/91)} & \multicolumn{2}{c|}{TinyImagenet (280/44)} & \multicolumn{2}{c}{Total (1350/232) } \\\cline{2-11}
                       & TPR/FPR  & Time   & TPR/FPR     & Time     & TPR/FPR  & Time    & TPR/FPR   & Time    & TPR/FPR     & Time   \\\hline
NC\cite{wang2019neural}&\textbf{100/0}&230.7s&52.78/18.18&376.7s&54.16/12.82&4,229.9s&61.07/15.91&$>$24h& 47.19/15.09&-\\
FeatureRE\cite{wang2022rethinking}&75.00/0.00&281.6s&27.62/11.69&354.5s& 38.93/12.09&2,676.3s&69.29/15.91 &50,835s&44.0/11.64 & 12,058s\\
DF-TND\cite{wang2020practical}&40.00/50.00&24.9s&20.0/4.54&91.7s&7.08/8.97& 990.4s&0.7/2.3& 40,706s&7.33/9.05&8,937s\\
DQ \cite{fields2021trojan}& 15.00/45.00&\textbf{0.14ms}& 52.38/7.58&\textbf{0.18ms}& 65.24/21.98&\textbf{0.2ms}& 42.5/4.55&\textbf{0.37ms}&   57.78/15.51&\textbf{0.22ms}\\
FreeEagle\cite{fu2023freeeagle}&35.00/10.00 & 6.36s&25.71/16.89 & 65.6s&43.81/12.09 & 274.0s&15.7/6.82&1,778.1s & 35.04/12.5  & 537.5s\\
MM-BD\cite{wang2024mm} & 60.00/22.7& 2.0s&31.90/25.97 & 15.9s&40.36/15.38 & 64.7s &20.0/15.91&1,810.1s &35.1/19.4&411.7s \\
BARBIE\cite{zhang2025barbie}\tnote{$\dagger$} & 95.00/15.00 & 91.0s  &86.19/62.33  & 176.7s &86.67/61.54  &3,626.1s   &92.14/68.18  &19,235s & 87.85/59.05 &6107.7s\\\hline
% \textbf{DFBScanner}& \textbf{100/0} & \underline{1.17ms} &\textbf{88.33/1.52} &\underline{1.14ms} &\textbf{99.03/0}& \underline{0.995ms}&\textbf{97.86}/2.27&\underline{1.13ms}&\textbf{97.17/0.95}&\underline{1.06ms}\\
\textbf{HTell}&\textbf{100/0} &\underline{1.26ms} &\textbf{98.10/3.90} &\underline{8.46ms}&\textbf{99.05/1.10} &\underline{12.98ms} &\textbf{99.29/0} &\underline{15.29ms}&\textbf{98.96/1.72} & \underline{12.34ms}\\\hline
&\multicolumn{2}{c|}{MNIST (20/20)}&\multicolumn{2}{c|}{CIFAR10 (700/147)}&\multicolumn{2}{c|}{GTSRB(3010/175)}& \multicolumn{2}{c|}{TinyImagenet (1120/84)}& \multicolumn{2}{c}{Total(4850/426)}\\\hline
% \textbf{DFBScanner}&100/0&\underline{1.17ms}&\textbf{90.16/7.14}&\underline{1.14ms}&\textbf{98.41/0.67}&\underline{0.995ms} &\textbf{98.21/7.14} &\underline{1.13ms}&\textbf{97.22/4.19}&\underline{1.05ms}\\
\textbf{HTell}&\textbf{100/0} &\underline{1.26ms} &\textbf{97.43/2.72} &\underline{8.54ms}&\textbf{99.27/1.71} &\underline{12.95ms} &\textbf{99.38/2.38} &\underline{15.31ms}&\textbf{99.03/2.11} & \underline{12.69ms}\\\bottomrule
%97.39/1.59
\end{tabular}
\begin{tablenotes}
  \footnotesize
  \item[$\dagger$] denotes that it can only identify whether the model is backdoored. So, its TPR are not as accurate as other methods.
\end{tablenotes}
\end{threeparttable}
\vspace{-4mm}
\end{table*}

% Detection success rate: TPR (true positive rate of detecting backdoor models) $\times$ ACC (accuracy of identifying backdoor labels)

\subsection{Backdoor Detection}

%720/ 7 120 714The number of benign and backdoor models of MNIST, CIFAR10, GTSRB, and TinyImageNet are 20/20, 210/77, 840/91, and 280/44, respectively.

\subsubsection{Backdoor Detection Performance}
\label{sub:performance}

% We evaluated HTell both in this small and full benchmark.
% To determine configurations in FreeEagle, BARBIE and HTell, the used models are not included in this two benchmarks.

Table \ref{tab:comparsion} demonstrates the detection accuracy and time cost of different methods. Since NC is designed primarily for patch-based backdoor attacks, it shows a high performance for BadNet, but fails to detect other complex backdoor attacks. NC and FeatureRE require the training dataset for trigger reversion, resulting in substantial time overhead. DQ detects backdoors by only identifying the maximum average weights in the final layer, thereby achieving the fastest detection speed. However, its detection performance is not robust. For example, DQ can achieve 100/9.09\% and 100/0\% TPR/FPR for GNet on CIFAR10 and GTSRB, respectively, but fails to detect attacks against PR18, VIT on TinyImageNet, and models on MNIST. DF-TND, FreeEagle, MM-BD, and BARBIE all use random noise as input and perform gradient optimization algorithms to reverse triggers or backdoor anomalies. MM-BD is designed specifically to detect backdoor attacks with arbitrary backdoor patterns. MM-BD outperforms DF-TND, but shows comparable efficacy to FreeEagle. BARBIE achieves a high TPR for backdoor models, but its FPR is also extremely high. The main reason may be that our benchmark contains diverse clean models trained from scratch with different seeds \cite{somepalli2022can}, leading BARBIE's metric to be less robust.
% Besides, BARBIE requires 60 clean models per architecture-dataset pair to determine its configuration
% These experimental results reveal 2 critical limitations in existing approaches: \textit{low robustness} and \textit{high computational costs}. While they achieve near-perfect accuracy under specific experimental conditions (\textit{i.e.}, fixed dataset-model-attack combinations), their performance deteriorates drastically, even catastrophically, when these precise conditions are altered.

% However, this Trojan signature is not stable due to differences in dataset distribution, model structures, attack trigger complexity, and attack implantation methods. Hence, the detection performance of DQ is still not satisfactory for the tasks of detecting our diverse backdoor attacks.
% However, MM-BD was only evaluated on a few model architectures (e.g., ResNet, VGG, and MobileNet), resulting in limited stability of its conclusions with our benchmark.

We evaluate HTell both on the small and full benchmark. HTell achieves a higher detection performance than other methods both in terms of detection accuracy and detection time cost. NC, DF-TND, MM-BD, FeatureRE, and BARBIE incur substantial time costs due to their requirement for iterative RE of triggers or anomalies. They must perform multiple RE iterations across either the full model or the backbone for each class to identify optimal triggers or anomalies. Consequently, these methods suffer from: (a) prohibitive backpropagation costs and (b) exponential redundancy from per-class RE computations. For example, NC requires more than 24 hours to analyze a ViT-16-b model trained on TinyImageNet, which is not acceptable in real-world scenarios. FreeEagle and BARBIE shift gradient-based analysis to the tail sub-network, thereby reducing the time cost of backpropagation. But they still require performing their analysis per class. BARBIE takes over 5 hours to analyze PreactResnet18 models trained on TinyImageNet. MM-BD employs gradient ascent to estimate the maximum margin of logits per class, thereby significantly reducing the time cost. Nevertheless, MM-BD still takes at least one and a half hours to analyze a ViT-16-b model.

%DQ has the lowest detection time. But the Trojan signature it used is not very stable due to the complex weight update ways caused by different attacks.

HTell focuses on exposing backdoor signatures by inputting a batch of backdoor probes into the model's head at once and analyzing logit anomalies across all classes. Compared to existing methods, it does not require any backpropagation operations, nor does it perform separate operations for each class. Hence, it effectively avoids these time costs in (a) and (b). Table \ref{tab:comparsion} demonstrates that HTell achieves a near-perfect and stable accuracy (99.03\% TPR and 2.11\% FPR) across different datasets, higher than baselines, and its average detection time is much lower than existing methods, besides DQ. Specifically, HTell obtains an average detection time of 12.69 ms on a 3090 GPU, significantly less than existing methods. We also evaluate HTell using the small benchmark on three other platforms with different computation capabilities, including Intel i9-12900K CPU, Apple M2 CPU, and Raspberry Pi 5b. Different platforms do not affect the TPR and FPR of HTell since it only relies on the deterministic model forward. So, Table \ref{tab:timecost} only shows HTell's average detection time costs across different platforms. We can find that HTell is very fast and lightweight for more practical scenarios.

% deploy in scenarios with various computation sources.

\renewcommand{\arraystretch}{1.0}
\begin{table}[t!]
\footnotesize
  \caption{The average detection time across different platforms.}
  \label{tab:timecost}
  \scalebox{0.93}{
  \begin{tabular}{ccccc}
    \toprule
    Platform & 3090 GPU & Intel i9 CPU & Apple M2 & Raspberry Pi 5b\\\midrule
    Time (ms) & 12.34$\pm$0.2 &14.47$\pm$0.9 &73.07$\pm$1.3 &83.24$\pm$2.0 \\\bottomrule
  \end{tabular}
  }
  \vspace{-1mm}
\end{table}

In summary, through this large-scale evaluation, we demonstrate that by probing the head using random noise, HTell achieves a highly accurate and efficient data-free backdoor detection. HTell achieves $>$99\% detection accuracy while reducing detection time costs by $>$30,000× compared to gradient-based methods.

\subsubsection{Detecting Attacks with Diverse Patch Triggers}
\textbf{Basic patch attacks}: We consider backdoor attacks with 8 different patch triggers in Fig. \ref{fig:trigger} to evaluate HTell, which have different patch sizes, locations, and color perturbations. We train backdoor models using V16, ENet, GNet, R18, and PR18 on CIFAR10 and GTSRB with 4 (3, 5, 7, and 9) and 5 target classes (8, 16, 24, 32, and 40), respectively. So, on CIFAR10 and GTSRB, we conduct 32 and 40 backdoor models per architecture, respectively, and 20 clean models per architecture. In total, there are 360 backdoor models and 200 clean models as the patch test benchmark. We choose DQ, MM-BD, and BARBIE as baselines. These methods are all configured as used in Table \ref{tab:comparsion}.

Table \ref{tab:badnet} demonstrates the results. DQ achieves a great TPR and FPR for GNet (100\% TPR and $<$10\% FPR) and R18 models ($>$96.88\% TPR and $<$10\% FPR), but fails to V16 and PR18 (0\% TPR). Due to differences in terms of trigger size, color, and location, different patches may lead to various marginal distribution changes. So, MM-BD's accuracy is also not robust for different patches. For example, except for GNet, MM-BD can successfully detect all backdoor models of the other 4 architectures on CIFAR10 for the patch \textbf{t4}, but MM-BD's TPR is only 4/20 for \textbf{t8}. BARBIE still suffers from a high FPR. HTell is robust on this benchmark and achieves $>$95.0\% TPR and $<$5\% FPR across all different triggers.

\renewcommand{\arraystretch}{1.0}
\begin{table}[t!]
\centering
\footnotesize
\caption{The TPR/FPR (\%) of detecting backdoor attacks with various patch triggers. C and G denote CIFAR10 and GTSRB datasets, respectively.}
\label{tab:badnet}
\scalebox{0.84}{
\begin{tabular}{c|c|ccccc}
\toprule
Method &  & ENet & V16 & GNet & R18 & PR18 \\\hline
DQ & \multirow{4}{*}{C} &59.38/10.0 &3.13/0.0 &\textbf{100}/10.0&\textbf{96.88/0.0}&0.0/10.0 \\
MM-BD &  &87.5/15.0 &18.75/40.0 &6.25/50.0 &25.0/10.0 &46.88/20.0 \\
BARBIE & &100/40.0 &96.88/5.0 &34.38/80.0 &100/30.0 &43.75/60.0 \\
HTell & &\textbf{96.88/0.0}  &\textbf{100/5.0}&\textbf{100/0.0} &\textbf{96.88/0.0}&\textbf{96.88/5.0} \\
\hline
DQ &  \multirow{4}{*}{G}&77.5/5.0  &45.0/10.0 &\textbf{100/0.0}&100/10.0&0.0/10.0 \\
MM-BD &  &42.5/20.0 &85.0/15.0 &17.5/10.0 &77.5/5.0 &\textbf{95.0}/35.0  \\
BARBIE & &100/80.0 &97.5/80.0 &100/75.0 &97.5/45.0 &100/45.0 \\
HTell &  &\textbf{97.5/0.0}  & \textbf{100/5.0} &\textbf{100/0.0} & \textbf{100/0.0}&\textbf{95.0/0.0} \\
\bottomrule
\end{tabular}
}
% \vspace{-4mm}
\end{table}

\textbf{Adaptive patch attacks}: We further study two adaptive patch attacks: Adaptive-patch \cite{qi2023adappatch} and LOTUS \cite{cheng2024lotus}. Adaptive-patch \cite{qi2023adappatch} can generate adaptive patches to avoid latent separability, making benign and backdoor latent features close in the latent space. LOTUS partitions poison samples into multiple groups and implants different triggers in each partition to enhance attack resilience. Those two attacks further challenge backdoor detection compared to existing patch attacks. We follow their default settings to train backdoor models. Specifically, we train 44 backdoor models of ResNet20 and V16, and R18 and VGG11 on CIFAR10 (all 10 classes) and GTSRB (12 different classes) via the adaptive-patch and LOTUS attack, respectively. In total, there are 88 backdoor models and 88 clean models in this adaptive patch test benchmark. We also configure these detection methods as used in Table \ref{tab:comparsion}.

Table \ref{tab:advancepatch} demonstrates the detection results against adaptive patch attacks. DQ detects Adaptive-patch attacks accurately, but fails to detect LOTUS. LOTUS stamps different triggers in different partitions of poisoned samples, leading to complex backdoor perturbations. Analyzing only the final-layer average weight is not sufficient for backdoor detection. MM-BD fails to detect both two adaptive patch attacks. BARBIE can detect most backdoor models, but its high FPR makes its accuracy not convincing. HTell totally identifies 87 backdoor models with their correct backdoor targets among 88 backdoor models, while achieving a low FPR of 3/88.

\renewcommand{\arraystretch}{1.0}
\begin{table}[t!]
\centering
\footnotesize
\caption{The TPR/FPR of detecting adaptive patch attacks.}
\label{tab:advancepatch}
\scalebox{0.79}{
\begin{tabular}{c|c|cccc|cccc}
\toprule
\multirow{3}{*}{Method} & \multirow{3}{*}{} & \multicolumn{4}{c|}{LOTUS}& \multicolumn{4}{c}{Adaptive-patch}\\\cline{3-10}
 &  &\multicolumn{2}{c}{R18} &\multicolumn{2}{c|}{V11} & \multicolumn{2}{c}{R20} & \multicolumn{2}{c}{V16}\\\cline{3-10}
 &  &TPR$\uparrow$ &FPR$\downarrow$ &TPR$\uparrow$ &FPR$\downarrow$ &TPR$\uparrow$ &FPR$\downarrow$ &TPR$\uparrow$ &FPR$\downarrow$\\\hline
DQ  & \multirow{3}{*}{C} &1/10 &\textbf{0/10} &1/10 &\textbf{0/10} &\textbf{10/10} &\textbf{0/10} &\textbf{10/10}&\textbf{0/10}\\
MM-BD &                         &0/10 &\textbf{0/10} &0/10 &1/10 &2/10 &1/10 &0/10 &4/10 \\
BARBIE &                        &\textbf{10/10} &8/10 &\textbf{10/10} &7/10 &\textbf{10/10} &1/10 &9/10 &3/10  \\
HTell & &\textbf{10/10}&\textbf{0/10}&\textbf{10/10}&1/10 &\textbf{10/10} &\textbf{0/10} &\textbf{10/10}&1/10\\
\hline
DQ & \multirow{3}{*}{G} &3/12 &\textbf{0/12} &1/12 &\textbf{0/12} &11/12&\textbf{0/12} & \textbf{12/12}&2/12\\
MM-BD &  &0/12 & \textbf{0/12}&0/12&1/12  & 3/12&1/12 & 0/12 &2/12\\
BARBIE &                        &\textbf{12/12} &7/12 &\textbf{12/12} &5/12 &\textbf{12/12} &4/12 &\textbf{12/12} &6/12  \\
HTell &  &\textbf{12/12} &\textbf{0/12} &11/12 &1/12 &\textbf{12/12} &\textbf{0/12} &\textbf{12/12} &\textbf{0/12}\\\bottomrule
\end{tabular}
}
% \vspace{-4mm}
\end{table}

\subsubsection{Impact of Network Sizes and Poisoning Rates}
Poisoning rates $\lambda$ and network sizes may also result in diverse backdoor anomalies. So, we train backdoor models on CIFAR10 using these 8 patch triggers in Fig. \ref{fig:trigger} with two backdoor targets 3 and 9 in 3 different $\lambda$ (including 0.01, 0.05, and 0.1) and 3 different network sizes (including ResNet18, ResNet34, and ResNet50). For each network architecture, we train 20 clean models from scratch with different seeds. In total, we conduct a benchmark, including 144 backdoor models and 60 corresponding clean models, to evaluate DQ, MM-BD, BARBIE, and HTell. We further train another 30 clean and benign models to configure these methods.

% to determine these methods' detection configurations.

Table \ref{tab:networksize} demonstrates the detection accuracy and time cost of these methods. In terms of poisoning rate, a smaller $\lambda$ can affect the accuracy of these methods. For example, the TPRs of DQ for backdoor models under $\lambda=0.1$ are all higher than those for backdoor models under other small $\lambda$ of 3 different network architectures. HTell's TPR under $\lambda=0.01$ is also lower than the one under $\lambda=0.1$. But HTell still identifies most backdoor models correctly. In terms of network sizes, BARBIE's detection accuracy gradually decreases as the network size increases. We keep the same iteration number for BARBIE's gradient-based anomaly identification across different network architectures. It may require more iterations to improve BARBIE's performance for larger networks, which, however, will result in more detection time costs. HTell's detection accuracy is almost unaffected by network size (93.75\% TPR for ResNet18, 100\% TPR for ResNet34, and 91.67\% TPR for ResNet50). HTell's detection time slightly changes since the head dimensions $D$ of ResNet50 are 2048, larger than ResNet18 and ResNet34 (their $D$ are all 512).

\renewcommand{\arraystretch}{1.0}
\begin{table}[t]
\centering
\footnotesize
\caption{The TPR/FPR of detecting backdoor attacks with various patch triggers under different network sizes and poisoning rates.}
\label{tab:networksize}
\scalebox{0.9}{
\begin{tabular}{c|c|ccccc}
\toprule
\multirow{2}{*}{Network} & \multirow{2}{*}{Method}  & $\lambda$=0.01 & $\lambda$=0.05 & $\lambda$=0.1 &\multirow{2}{*}{FPR$\downarrow$} & \multirow{2}{*}{Avg. Time}\\\cline{3-5}
 &&TPR$\uparrow$ &TPR$\uparrow$ &TPR$\uparrow$ &&\\\hline
\multirow{4}{*}{R18}&DQ &11/16&13/16&13/16 &2/20&\textbf{0.16$\pm$0.02ms}\\
&MM-BD &5/16&4/16&1/16&3/20&28.62$\pm$1.8s\\
&BARBIE &\textbf{14/16}&14/16&\textbf{16/16}&6/20&67.2$\pm$3.1s\\
&HTell &\textbf{14/16}& \textbf{15/16}&\textbf{16/16}&\textbf{0/20}&\underline{2.21$\pm$0.2ms}\\
\hline
\multirow{4}{*}{R34}&DQ &11/16&15/16&\textbf{16/16}&4/20&\textbf{0.16$\pm$0.02ms}\\
&MM-BD &1/16&3/16&6/16&3/20&38.93$\pm$1.9s\\
&BARBIE &9/16&7/16&6/16&5/20&107.7$\pm$3.9s\\
&HTell &\textbf{16/16}&\textbf{16/16}&\textbf{16/16}&\textbf{1/20}&\underline{2.23$\pm$0.2ms}\\
\hline
\multirow{4}{*}{R50}&DQ &10/16&12/16&14/16&3/20&\textbf{0.2$\pm$0.02ms}\\
&MM-BD &3/16&11/16&9/16&19/20&51.34$\pm$2.5s\\
&BARBIE &6/16&5/16&2/16&7/20&148.7$\pm$5.5s\\
&HTell &\textbf{13/16}&\textbf{15/16}&\textbf{16/16}&\textbf{0/20}&\underline{8.18$\pm$0.3ms}\\
\bottomrule
\end{tabular}
}
% \vspace{-5mm}
\end{table}

\subsubsection{Detecting Parameter Modification Attacks}

We further study backdoor attacks injected via parameter modification attacks, including bit-flip \cite{rakin2020tbt, bai2022hardly} and direct parameter modification \cite{cao2024data}, which enable the online backdoor attacks. Different from data poisoning or training process manipulation, these attacks identify critical parameters and modify a few parts of parameter values to inject backdoor attacks while keeping original benign behaviors. For bit-flip attacks, we adopt two open-source bit-flip backdoor attacks (TBT \cite{rakin2020tbt} and HPT \cite{bai2022hardly}) to conduct backdoor models. Specifically, we follow their default setting to train 10 8-bit clean models of R18 and V16 on CIFAR10, respectively, and generate 20 8-bit backdoor models across all classes as the bit-flip test benchmark. We also train an additional 10 clean and 5 backdoor 8-bit models to determine detection configurations of different detection methods. DFBA \cite{cao2024data} is a data-free backdoor attack via parameter modification. We modify clean models of R18 and V16 on CIFAR10 and GTSRB via DFBA to generate backdoor models with backdoor targets of all 10 labels of CIFAR10 and 12 different labels of GTSRB as the data-free test benchmark (64 backdoor and 64 clean models). The detection configurations of DQ, MM-BD, BARBIE, and HTell follow as in Table \ref{tab:comparsion}.

Table \ref{tab:bitflip} shows the detection results for bit-flip backdoor attacks. BARBIE and HTell demonstrate superior detection accuracy over MM-BD and DQ. This performance advantage stems from fundamental properties of bit-flip attacks: as studied in \cite{rakin2020tbt, bai2022hardly}, flipping the final layer parameters constitutes the most efficient method for implanting backdoors with minimal bit modifications. Hence, these two attacks inherently expand the decision boundary of the backdoor target, amplifying the response to the latent feature of triggers. HTell achieves an average detection time of 2.53 ms for 8-bit models, better than MM-BD and BARBIE. Table \ref{tab:datafree} demonstrates the detection results against DFBA \cite{cao2024data}. DFBA searches a backdoor path from the input layer to the output layer to modify parameter values for backdoor injection. This modification may also result in an expanded decision boundary before the final head to ensure ASR. HTell achieves a TPR of 100\% and FPR of 0\%. In the aspect of detection time, only DQ and HTell can detect DFBA in time. These results demonstrate HTell's online backdoor detection capabilities for practical model auditing.

\begin{table}[t]
\centering
\footnotesize
\caption{Detection performance against bit-flip attacks.}
\label{tab:bitflip}
\begin{tabular}{c|cccc|c}
\toprule
\multirow{2}{*}{Method} & \multicolumn{2}{c}{ResNet18} & \multicolumn{2}{c|}{VGG16} & \multirow{2}{*}{Detection Time} \\\cline{2-5}
 &TPR$\uparrow$ & FPR$\downarrow$&TPR$\uparrow$ &FPR$\downarrow$&\\\hline
 DQ    &13/20&\textbf{0/10} &11/20&\textbf{0/10} & \textbf{0.12$\pm$0.02ms} \\
 MM-BD &7/20&1/10  &10/20&1/10 & 30.32$\pm$3.9s \\
 BARBIE &\textbf{20/20} &1/10  &\textbf{20/20} &2/10 & 125.7$\pm$3.2s\\
 HTell & \textbf{20/20}&\textbf{0/10} & 19/20&\textbf{0/10} &  \underline{2.53$\pm$0.1ms}\\\bottomrule
\end{tabular}
% \vspace{-4mm}
\end{table}

\begin{table}[t]
\centering
\footnotesize
\caption{Detection performance against the DFBA attack \cite{cao2024data}. The attack time is the average time of injecting a backdoor into a model via DFBA. The detection time is the average backdoor detection time of these compared detection methods.}
\label{tab:datafree}
\scalebox{0.88}{
\begin{tabular}{c|c|cccc|c|c}
\toprule
\multirow{2}{*}{Method}&\multirow{2}{*}{} & \multicolumn{2}{c}{ResNet18} & \multicolumn{2}{c|}{VGG16} & \multirow{2}{*}{\makecell{Attack \\ time}} & \multirow{2}{*}{\makecell{Detection\\time}} \\\cline{3-6}
 &&TPR$\uparrow$ & FPR$\downarrow$&TPR$\uparrow$ &FPR$\downarrow$& &\\\hline

DQ    & \multirow{4}{*}{C}&\textbf{10/10}  &\textbf{0/10}  &1/10 &\textbf{0/10} &\multirow{4}{*}{\makecell{10.7$\pm$ \\0.5ms}} &\textbf{0.13$\pm$0.02ms} \\
 MM-BD &&0/10&2/10   &1/20&4/10  & &28.92$\pm$1.9s \\
 BARBIE &&\textbf{10/10} &5/10  &\textbf{10/10} &\textbf{0/10} & &81.5$\pm$3.1s\\
 HTell && \textbf{10/10}&\textbf{0/10} & \textbf{10/10}&\textbf{0/10} &  &\underline{2.30$\pm$0.2ms}\\\hline
DQ    & \multirow{4}{*}{G}&\textbf{22/22} &\textbf{0/22}  &0/22 &2/22  &\multirow{4}{*}{\makecell{10.9$\pm$\\0.7ms}} &\textbf{0.13$\pm$0.02ms} \\
MM-BD  & & 0/22&1/16   &1/22 &3/16 & & 59.1$\pm$3.2s \\
 BARBIE  &&\textbf{22/22} &10/22  &\textbf{22/22} &14/22 & &897.9$\pm$8.1s\\
 HTell  && \textbf{22/22}&\textbf{0/22} & \textbf{22/22}&\textbf{0/22} &  &\underline{3.79$\pm$0.2ms}\\\hline
\end{tabular}
}
% \vspace{-2mm}
\end{table}

\subsubsection{Detecting Backdoor Attacks on SSL}

We choose two backdoor attacks (BadEncoder \cite{jia2022badencoder} and DRUPE \cite{tao2024distribution}) against SSL to further evaluate HTell. Backdoor attacks on SSL construct a backdoored image encoder on a pre-training dataset, which will be provided for different downstream tasks. In this scenario, the downstream users utilize this pre-trained encoder to extract latent features from their specific downstream datasets and train a downstream head into transform latent features to downstream prediction labels. These downstream classifiers inherit the backdoor behavior from the backdoor encoder. These two attacks minimize the distance of latent representations between backdoor and clean samples in their backdoor encoders, thereby ensuring high stealth.

Similar to \cite{zhang2025barbie}, we adopt CIFAR10 as the pre-training dataset and employ ResNet18 to train the encoder via SimCLR \cite{chen2020simple} (a popular SSL algorithm). For downstream tasks, we adopt two downstream datasets (GTSRB and SVHN \cite{netzer2011reading}) and use an FC neural network with two hidden layers to train downstream classifiers. We train 40 backdoor classifiers of BadEncoder and DRUPE, respectively, and 40 clean classifiers on each downstream datasets. We use 60 backdoor and 60 clean classifiers as an SSL test benchmark, and the remaining classifiers are used for determining detection configurations of detection methods. We use normal distribution $\mathcal{N}(0, 0.1)$ to generate backdoor probes, and use $\mathbf{r}^{\text{mean}}$ as the detection indicator with a threshold $\tau$=0.179 and $\tau$=0.081 for SVHN and GTSRB, respectively. For BARBIE, we train another 40 clean classifiers to determine its detection configuration.

Table \ref{tab:ssl} demonstrates detection accuracy and time costs of different methods against SSL backdoor attacks. DQ and MM-BD fail to detect most SSL backdoor models. BARBIE achieves an acceptable detection TPR and FPR, especially for the BadEncoder attack. HTell achieves a TPR of 99.17\% and FPR of 0\% for two SSL backdoor attacks. DRUPE employs the sliced-Wasserstein distance between latent representations of backdoor and clean samples to enable a more stealthy attack. But it can still cause an abnormal decision boundary to the backdoor target. Hence, HTell can accurately detect these backdoor attacks.

\renewcommand{\arraystretch}{1.0}
\begin{table}[t]
\centering
\footnotesize
\caption{Detection performance against backdoor attacks on SSL. The used clean and backdoor encoders are all trained on CIFAR10.}
\label{tab:ssl}
\scalebox{0.9}{
\begin{tabular}{c|c|ccc|c}
\toprule
\multirow{2}{*}{Method}&\multirow{2}{1.3cm}{\centering Downstream Dataset} & BadEncoder & DRUPE &\multirow{2}{*}{FPR$\downarrow$} & \multirow{2}{1.2cm}{\centering Detection Time} \\\cline{3-4}
 &&TPR$\uparrow$ &TPR$\uparrow$ &\\\hline

DQ    & \multirow{4}{*}{SVHN}&0/30 &3/30 &\textbf{0/30}  &\textbf{0.11$\pm$0.02ms} \\
 MM-BD &&3/30   &3/30&6/30  &31.22$\pm$1.7s \\
 BARBIE &&27/30  &21/30 &3/30  &87.7$\pm$3.2s\\
 HTell && \textbf{30/30} & \textbf{30/30}&\textbf{0/30}  &\underline{3.02$\pm$0.6ms}\\\hline

DQ    & \multirow{4}{*}{GTSRB}&0/30 &0/30 &\textbf{0/30}  &\textbf{0.12$\pm$0.02ms} \\
MM-BD  & & 0/30   &1/30 &1/30  & 62.37$\pm$2.1s \\
 BARBIE  &&\textbf{30/30} &\textbf{30/30} &3/30  &954.2$\pm$8.9s\\
 HTell  && 29/30 & \textbf{30/30}&\textbf{0/30}   &\underline{5.51$\pm$0.6ms}\\\bottomrule
\end{tabular}
}
% \vspace{-5mm}
\end{table}

% \renewcommand{\arraystretch}{1.0}
% \begin{table}[t]
% \centering
% \footnotesize
% \caption{Detection performance against backdoor attacks on SSL. The used clean and backdoor encoders are all trained on CIFAR10.}
% \label{tab:ssl}
% \scalebox{0.9}{
% \begin{tabular}{c|c|cccc|c}
% \toprule
% \multirow{2}{*}{Method}&\multirow{2}{1.3cm}{\centering Downstream Dataset} & \multicolumn{2}{c}{BadEncoder} & \multicolumn{2}{c|}{DRUPE} & \multirow{2}{1.2cm}{\centering Detection Time} \\\cline{3-6}
%  &&TPR$\uparrow$ & FPR$\downarrow$&TPR$\uparrow$ &FPR$\downarrow$ &\\\hline
%
% DQ    & \multirow{4}{*}{SVHN}&0/30  &\textbf{0/30}  &3/30 &\textbf{0/30}  &\textbf{0.11$\pm$0.02ms} \\
%  MM-BD &&3/30&6/30   &3/30&6/30  &31.22$\pm$1.7s \\
%  BARBIE &&27/30 &3/30  &21/30 &3/30  &87.7$\pm$3.2s\\
%  HTell && \textbf{30/30}&\textbf{0/30} & \textbf{30/30}&\textbf{0/30}  &\underline{3.02$\pm$0.6ms}\\\hline
%
% DQ    & \multirow{4}{*}{GTSRB}&0/30 &\textbf{0/30}  &0/30 &\textbf{0/30}  &\textbf{0.12$\pm$0.02ms} \\
% MM-BD  & & 0/30&1/30   &1/30 &1/30  & 62.37$\pm$2.1s \\
%  BARBIE  &&\textbf{30/30} &3/30 &\textbf{30/30} &3/30  &954.2$\pm$8.9s\\
%  HTell  && 29/30&\textbf{0/30} & \textbf{30/30}&\textbf{0/30}   &\underline{5.51$\pm$0.6ms}\\\bottomrule
% \end{tabular}
% }
% \vspace{-5mm}
% \end{table}
% BARBIE: vgg  cifar10 1/0
%resnet cifar10 1/0.2381
%resnet gtsrb 1/0.733
%vgg gtsrb 1/0.8421
% \subsubsection{Backdoor Detection of Bit-Flip Attacks}

\subsubsection{Detecting Clean-label Backdoor Attacks}
In this section, we evaluated HTell with 4 clean-label backdoor attacks: SIG \cite{mauro2019sig}, LC \cite{turner2019labelconsistent}, Narcissus \cite{zeng2023narcissus}, and DataFree \cite{lv2023data}. Different from the above dirty-label backdoor attacks, those clean-label backdoor attacks do not change the backdoor target label. SIG employs a sinusoidal signal as the trigger and attacks using the model's sensitivity to high-frequency signals. LC generates perturbations through optimization algorithms to make the middle-layer features closer to the average features of the target class. Narcissus employs a pre-trained surrogate model to generate a unique trigger for each sample that is highly relevant to its content. DataFree fine-tunes the original model into the backdoored one with a substitute dataset. Clean-label attacks tend to be more stealthy.

Since clean-label attacks may not induce a unique target-response peak, the original HTell target-identification rule is not directly applicable. We therefore evaluate a lightweight variant, HTell-SIM, which compares the response profile of a queried model with the average profile of calibrated clean models. HTell-SIM is used only to answer RQ1 in this setting. Specifically, inspired by \cite{zhang2025barbie}, we compare the differences of decision boundaries between backdoor and clean models to detect backdoor models. With the statistical indicator $\mathbf{r}$ of a queried model $\mathcal{M}$, HTell-SIM detects backdoors as follows:
\begin{equation}\label{equ:sim}
  \text{cosine}(\mathbf{r}, \overline{\mathbf{r}}^{\text{clean}})>\tau^{\text{all2all}},
\end{equation}
where $\overline{\mathbf{r}}^{\text{clean}}$ is the average indicator of a few of clean models and $\tau^{\text{all2all}}$ is a threshold to distinguish clean and backdoor models. Similarly, we also extend DQ \cite{fields2021trojan} with the above metric as DQ-SIM to enable detection in this scenario. We train 10 and 12 backdoor models of ResNet18 on CIFAR10 (all labels) and GTSRB (12 different labels), respectively, for each attack. In total, we conduct a clean-label test benchmark with 88 backdoor models and 30 clean models. We also train another 24 backdoor and 10 clean models to determine configurations of DQ-SIM, BARBIE, and HTell-SIM.

Table \ref{tab:cleanlabel} demonstrates the detection accuracy of DQ-SIM, BARBIE, and HTell-SIM. Note that these three methods can only answer \textbf{RQ1}. DQ-SIM is valid for Narcissus attacks on CIFAR10 and LC, Narcissus, and DataFree on GTSRB. BARBIE and HTell-SIM achieve similar TPR, \textit{i.e.}, 96.59\% and 97.73\%, respectively. But, HTell-SIM can achieve an FPR of 10\% lower than BARBIE (30\%).
% CIFAR10 normal l2, $\tau$=6.19e-5
%
% GTSRB normal l2, $\tau$=0.00112

\renewcommand{\arraystretch}{1.0}
\begin{table}[t!]
\centering
\footnotesize
\caption{The TPR/FPR of detecting clean-label backdoor attacks. -SIM means a variant extended via Equ. (\ref{equ:sim}).}
\label{tab:cleanlabel}
\scalebox{1.0}{
\begin{tabular}{c|c|ccccc}
\toprule

&\multirow{2}{*}{Method}& SIG & LC & Narcissus & DataFree & \multirow{2}{*}{FPR$\downarrow$}\\\cline{3-6}
& &TPR$\uparrow$ &TPR$\uparrow$ &TPR$\uparrow$ &TPR$\uparrow$ &\\\hline
\multirow{3}{*}{C} & DQ-SIM &5/10 &3/10 &9/10 &1/10 &4/15 \\
& BARBIE &\textbf{10/10} &9/10 &9/10 &\textbf{10/10}  &5/15\\
& HTell-SIM &9/10 &\textbf{10/10} &\textbf{10/10} &\textbf{10/10} &\textbf{2/15} \\\hline
\multirow{3}{*}{G} & DQ-SIM &7/12 &10/12 &\textbf{12/12} &\textbf{12/12} &4/15 \\
& BARBIE &\textbf{12/12} &\textbf{11/12} &\textbf{12/12} &\textbf{12/12} & 4/15\\
& HTell-SIM &\textbf{12/12} &\textbf{11/12} &\textbf{12/12} &\textbf{12/12} &\textbf{1/15} \\\bottomrule
\end{tabular}
}
% \vspace{-2mm}
\end{table}

\renewcommand{\arraystretch}{1.0}
\begin{table}[t!]
\centering
\footnotesize
\caption{The TPR/FPR of detecting all-to-all backdoor attacks.}
\label{tab:all2all}
\scalebox{1.0}{
\begin{tabular}{c|c|cccccc}
\hline
&\multirow{2}{*}{Model}&\multicolumn{2}{c}{DQ-SIM} & \multicolumn{2}{c}{BARBIE} & \multicolumn{2}{c}{HTell-SIM}\\\cline{3-8}
&&TPR$\uparrow$&FPR$\downarrow$&TPR$\uparrow$&FPR$\downarrow$&TPR$\uparrow$&FPR$\downarrow$\\\hline
\multirow{7}{*}{C} & ENb3 & 45/50 &7/15 & 49/50 & 12/15 &\textbf{50/50} &\textbf{1/15} \\
&R18 &31/50 & 6/15 &\textbf{50/50} &5/15 &\textbf{50/50} & \textbf{0/15} \\
&PR18 &40/50 &7/15 &25/50 &10/15 &\textbf{45/50} &\textbf{0/15} \\
&V16 &2/50 & 2/15 &47/50 &8/15 &\textbf{50/50} &\textbf{0/15} \\
&CNN6 &48/50 &\textbf{3/15} &44/50 &\textbf{3/15} &\textbf{50/50} &\textbf{3/15} \\
&GNet &25/50 & 7/15 &34/50 &12/15 &\textbf{48/50} &\textbf{3/15} \\
&SNet &49/50 & 5/15 &45/50 &13/15 &\textbf{50/50} &\textbf{0/15} \\\hline
\multirow{7}{*}{G} & ENb3 & 48/50 &4/15 & 39/50 & 9/15 &\textbf{50/50} &\textbf{0/15} \\
&R18 &\textbf{50/50} & 4/15 &\textbf{50/50} &5/15 &\textbf{50/50} & \textbf{0/15} \\
&PR18 &\textbf{50/50} &6/15 &25/50 &10/15 &\textbf{50/50} &\textbf{0/15} \\
&V16 &46/50 & 1/15 &\textbf{50/50} &8/15 &\textbf{50/50} &\textbf{0/15} \\
&CNN6 & 46/50 &7/15 &\textbf{50/50} &10/15 &\textbf{50/50} &\textbf{1/15} \\
&GNet &26/50 & 7/15 &48/50 &12/15 &\textbf{50/50} &\textbf{0/15}  \\
&SNet &44/50 & \textbf{0/15} &36/50 &3/15 & \textbf{50/50} &\textbf{0/15} \\\hline
\end{tabular}
}
\vspace{-2mm}
\end{table}

\subsubsection{Detecting All-to-All Backdoor Attacks}

In all‐to‐all backdoor attacks, poisoned samples coming from different source classes have different target labels. These attacks also may not induce a unique target-response peak. Hence, we use the variant HTell-SIM to detect all-to-all backdoors. We train 5 all-to-all backdoor models for each model-attack pair on CIFAR10 and GTSRB (7 network architectures and 10 backdoor attacks), as used in Table \ref{tab:comparsion}. In an all-to-all  backdoor model, we follow a loop permutation to generate target and source class pairs, \textit{i.e.}, each class $k\in K$ is a poison target class with a source class $(k+1)\ \text{mod}\ K$. In total, we conduct 700 all-to-all backdoor models as the all-to-all test benchmark. We compare HTell-SIM with DQ-SIM and BARBIE to inspect these models and use another 70 clean and backdoor models to determine their detection configurations.

Table \ref{tab:all2all} demonstrates the detection results. Any modification to the dataset or model training process may lead to changes in the model decision space. This is even more obvious in all-to-all attacks. All these 3 methods focus on the anomalies inside the model's head. HTell-SIM uses random probes to fill the decision space of all classes as much as possible. The probes' corresponding output logits provide a sketch of all decision spaces. This metric in Equ. (\ref{equ:sim}) demonstrates anomalies in decision spaces introduced by backdoor attacks significantly. HTell-SIM achieves the best detection accuracy among these 3 methods.

% But, it also fails to detect in some cases, mainly including the CIFAR10-PR18-LF pair.

%
% \renewcommand{\arraystretch}{1.0}
% \begin{table}[t!]
% \centering
% \footnotesize
% \caption{The TPR/FPR (\%) of detecting all-to-all backdoor attacks.}
% \label{tab:all2all}
% \scalebox{0.65}{
% \begin{tabular}{c|c|ccccccc}
% \hline
% Method& & ENb3 & R18 & PR18 &V16 &CNN6 &GNet &SNet\\\hline
% DQ-SIM & \multirow{3}{*}{C} &90.0/46.7 &62.0/40.0 &80.0/46.7 &4.0/13.3 &96.0/20.0 &50.0/46.7 &98.0/33.3\\
% BARBIE &&98.0/80.0 &100.0/33.3& 50.0/66.7 &94.0/37.5 &88.0/20.0 &68.0/80.0 &90.0/86.7 \\
% HTell &  &100.0/6.7 &100.0/0.0 &90.0/0.0 &100.0/0.0 &100/20.0 &96.0/20.0 &100.0/0.0\\\hline
% DQ-SIM & \multirow{3}{*}{G} &96.0/26.7 & 100.0/26.7&100.0/40.0 &92.0/6.7 &92.0/46.7 &52.0/46.7 &88.0/0.0\\
% BARBIE &  &78.0/60.0 &78.0/53.3 &96.0/46.7 &100.0/53.3 &100.0/66.7 &96.0/100.0 &72.0/20.0\\
% HTell &  & 100.0/0.0  &100.0/0 &100.0/0.0 &100.0/0.0 &100.0/6.7 &100.0/0.0 &100.0/0.0\\\hline
% \end{tabular}
% }
% \end{table}

\subsubsection{Detection with Poisoned Model Zoos}

HTell determines its detection threshold using a small configuration set of clean and backdoored models. However, in practical model-auditing scenarios, the defender may not have a fully trusted model set. For example, models collected from online may already contain a mixture of benign and backdoored models. In such a case, directly calibrating a fixed threshold from the collected models may be unreliable. To evaluate whether HTell can still provide useful detection signals without relying on a trusted threshold, we conduct a poisoned model zoo experiment.

% we evaluate whether HTell's score can rank backdoored models ahead of clean ones.

Table~\ref{tab:zoo} shows that HTell remains effective even when the model zoo itself is contaminated. On CIFAR10, HTell achieves high AP across all tested backdoor-to-clean ratios, with GNet and SNet reaching 100\% AP and ENb3/R18 consistently above 91\%. The mTPR also remains above 95\%, indicating that the detected response concentration usually corresponds to the true backdoor target. On GTSRB, HTell achieves nearly perfect AP and mTPR across all architectures and contamination ratios. These results demonstrate that HTell is not only fast under fixed-threshold data-free detection, but also useful as a threshold-free ranking signal in contaminated model repositories, enabling it to support practical model-zoo screening even without a fully trusted clean calibration set.

\renewcommand{\arraystretch}{1.0}
\begin{table}[t!]
\centering
\footnotesize
\caption{HTell's detection performance (\%) for poisoned model zoos.}
\label{tab:zoo}
\begin{adjustbox}{max width=\columnwidth}
\begin{tabular}{c|c|cccc|c}
\hline
Dataset &Backdoor&\multicolumn{4}{c|}{Average Precision} & \multirow{2}{*}{mTPR}\\\cline{3-6}
&/clean &ENb3  &R18 & GNet & \multicolumn{1}{c|}{SNet} & \\\hline
\multirow{4}{*}{CIFAR10}
&10/100  &94.98 &91.45  & 100.0 &100.0 & 95.92  \\
&40/100  &94.00 &95.00 & 100.0 &100.0  & 96.75 \\
&70/10   &96.80 &94.05 & 100.0 &100.0  & 97.36  \\
&100/100 &96.87 &95.29  & 100.0 &100.0  & 97.72\\\hline
\multirow{4}{*}{GTSRB}
&10/100  &100.0 &100.0  & 100.0 &100.0  & 100.0 \\
&40/100  &99.51 &100.0  & 100.0 &100.0  & 99.58  \\
&70/10   &100.0 &100.0  & 100.0 &100.0 & 100.0 \\
&100/100 &99.80 &100.0  & 100.0 &100.0  & 99.75\\\hline
\end{tabular}
\end{adjustbox}
\end{table}

\subsection{Possible Adaptive Attacks against HTell}

% (1) the adaptive attacker adds a regularization on parameters $\theta^c$ to the model training loss, similar to \cite{fields2021trojan}; (2) the adaptive attacker employs the classification part of benign models and freeze its parameters during backdoor implantation.
To analyze how an attacker can bypass HTell if she has the knowledge of HTell, we investigate two possible adaptive attacks against HTell:

\textbf{1) Adaptive Attack via Regularizing Parameters}. Similar to \cite{fields2021trojan}, the attack introduces an additional loss function to regularize $\theta^c$ as follows:
\begin{equation}
\mathcal{L}=\mathcal{L}_{\text{ce}}+\beta\sum_{i=1}^K\|\mathbf{w}_i-\overline{\mathbf{w}}\|^2+\gamma\sum_{i=1}^K(b_i-\overline{b})^2,
\end{equation}
where $\overline{\mathbf{w}}$ and $\overline{b}$ are the means of all classes of weights and biases, respectively, $\beta$ and $\gamma$ control the suppression intensity. This loss function forces $\mathbf{w}_l$ and $b_l$ of the backdoor class $l$ to stay close to other classes.

\textbf{2) Adaptive Attack via Freezing Parameters}. In this attack, the attacker first trains a clean model and then freezes its $\theta^c$ to retrain the model using the poison dataset. This method can defend against many backdoor detection methods (e.g., FreeEagle, DQ, and our HTell), but it may have an impact on the classification performance of benign samples since it can only be achieved by optimizing the backbone.

We train 140 backdoor models on CIFAR10 and GTSRB with 7 architectures and 10 attacks on 2 random backdoor targets per adaptive attack. In total, we conduct 560 backdoor models to evaluate HTell. Table \ref{tab:adaptive} illustrates these models' performance and HTell's detection accuracy. Both adaptive attacks can bypass HTell's detection to some extent. But the first attack can result in a higher decrease of BA than the second attack, and HTell can still detect part of backdoor models. These results indicate that HTell is vulnerable to head-constrained adaptive attacks. However, moving the probing point from the prediction head to the last feature block substantially recovers the detection rate, suggesting a practical mitigation direction. Note that this result may be further improved by reconfiguring HTell according to the hidden features in the last block of the backbone, rather than using the original configurations.

\renewcommand{\arraystretch}{1}
\begin{table}[t!]
\footnotesize
  \caption{HTell's detection TPR (\%) against adaptive attacks. * denotes HTell's detection results after moving the starting layer $L^C$ of $C$ to an upper layer $L^{C-1}$.}
    \label{tab:adaptive}
    \centering
    \scalebox{1}{
    \begin{tabular}{c|ccc|ccc}
    \hline
         \multirow{2}{*}{}&\multicolumn{3}{c|}{Parameter Regularization} &\multicolumn{3}{c}{Parameter Frozen}\\\cline{2-7}
         & $\Delta^{\text{BA}}/\Delta^{\text{ASR}}$ & TPR & TPR$^*$ & $\Delta^{\text{BA}}/\Delta^{\text{ASR}}$ & TPR & TPR$^*$\\\hline
         C &-1.36/-2.16 &53.6 &78.6 &-0.19/0.49 &10.7 & 92.1\\
         G  & -1.24/-2.57 &38.6 & 69.3 &-0.88/0.52&5.0 & 89.3\\\hline
    \end{tabular}
    }
    % \vspace{-5mm}
\end{table}

\subsection{Ablation and Sensitivity Analysis}

We further analyze key design choices of HTell as follows:

\subsubsection{Probe Distribution Analysis}

HTell employs either uniform or Gaussian probes according to the coarse latent activation range. To validate this design, we compare different probe distributions on representative architectures. As shown in Table~\ref{tab:probe_dist}, architectures with non-negative latent activations (e.g., ResNet and VGG) achieve better performance using uniform probes, while architectures with normalized or signed latent activations (e.g., EfficientNet and PreactResNet) benefit more from Gaussian probes. These results confirm that matching the probe distribution to the latent-space geometry improves the stability of head-level probing.

\begin{table}[t]
\centering
\caption{Impact of probe distributions.}
\label{tab:probe_dist}
\begin{tabular}{l|c|c}
\hline
Architecture & Uniform & Gaussian \\
\hline
ResNet18 & 99.1/1.1 & 81.3/8.2 \\
VGG16 & 98.7/0.0 & 75.0/12.5 \\
EfficientNet & 62.5/10.0 & 96.8/0.0 \\
PreactResNet18 & 57.1/9.1 & 98.1/1.1 \\
\hline
\end{tabular}
\end{table}

\subsubsection{Indicator Analysis}

We compare the two response indicators ($\mathbf{r}^{\text{l2}}$ and $\mathbf{r}^{\text{mean}}$) used in HTell with other two indicators: class-prediction proportion $\mathbf{r}^{\text{ratio}}=[\frac{1}{N}\sum_{n=1}^N\mathbb{I}(\hat{y}_n=i)]$ and class-maximum confidence $\mathbf{r}^{\text{max}}=[\max_{n=1}^Nz_{n,i}]$. Table \ref{tab:indicator} demonstrates their average detection performance. While $\mathbf{r}^{\text{ratio}}$ and $\mathbf{r}^{\text{max}}$ have specific advantages in certain model settings, $\mathbf{r}^{\text{mean}}$ performs best for all datasets at average and $\mathbf{r}^{\text{l2}}$ is more robust for PreactResNet18 due to its highly concentrated latent activations.
%This behavior suggests that different architectures may exhibit different forms of response concentration under backdoor implantation.

% We determine $\tau$ for four indicators using the configuration set and demonstrate their detection TPR and FPR on the full backdoor detection benchmark. We can see that due to differences in network architectures, these four indicators all have the capacity of identifying backdoor attacks to some extent, but their performance varies across different architectures and datasets. Among them, $\mathbf{r}^{\text{mean}}$ and $\mathbf{r}^{\text{l2}}$ show a better performance than others. In total, besides PreactResNet18, $\mathbf{r}^{\text{mean}}$ achieves a near-perfect and stable performance for most network architectures. Hence, we set $\mathbf{r}^{\text{mean}}$ as the detection indicator for most mdoels and only set $\mathbf{r}^{\text{l2}}$ for PreactResNet18. Note that $\mathbf{r}^{\text{ratio}}$ and $\mathbf{r}^{\text{max}}$ also have specific advantages in certain model settings. For example, $\mathbf{r}^{\text{ratio}}$ achieves an TPR of 99\% and FPR of 0\% for ENb3 on CIFAR10, better than $\mathbf{r}^{\text{mean}}$. Hence, given a small configuration set, the defender can determine which indicator should be used according to their detection accuracy.

\begin{table}[t]
\centering
\caption{Average TPR/FPR (\%) of HTell under different response indicators across datasets. The bold value is the best.}
\label{tab:indicator}
\begin{tabular}{l|c|c|c|c}
\hline
Dataset & $\mathbf{r}^{\text{ratio}}$ & $\mathbf{r}^{\text{max}}$ & $\mathbf{r}^{\text{mean}}$ & $\mathbf{r}^{\text{l2}}$ \\
\hline
CIFAR10 & 89.29/4.77 & 79.43/3.4 & \textbf{97.29/10.9} & 96.57/3.4 \\
GTSRB & 98.41/4.0 & 89.94/2.29 & \textbf{99.31/1.71} & 99.27/1.14 \\
TinyImageNet & 96.0/5.95 & 69.58/4.75 & \textbf{99.18/3.58} & 97.48/2.4 \\
MNIST & 100/10 & 90/45 & \textbf{100/0} & \textbf{100/0} \\
\hline
Overall & 96.49/4.9 & 83.71/13.0 & \textbf{98.99/5.2} & 98.47/2.1 \\
\hline
\end{tabular}
\end{table}

\subsubsection{Sensitivity to Gaussian Variance}

\begin{figure}[t!]
\centering
\begin{minipage}[c]{1\linewidth}
    \centering
    % \subfloat[SqueezeNet]{\includegraphics[width=0.49\linewidth]{figures/squeezenet_sigma_setting.pdf}}
    \subfloat[EfficientNet-B3]{\includegraphics[width=0.49\linewidth]{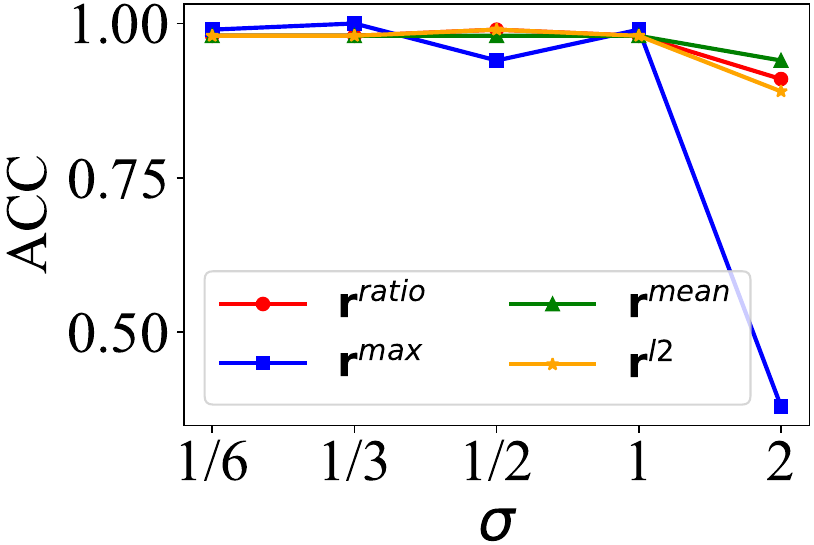}}
    \subfloat[PreactResNet18]{\includegraphics[width=0.49\linewidth]{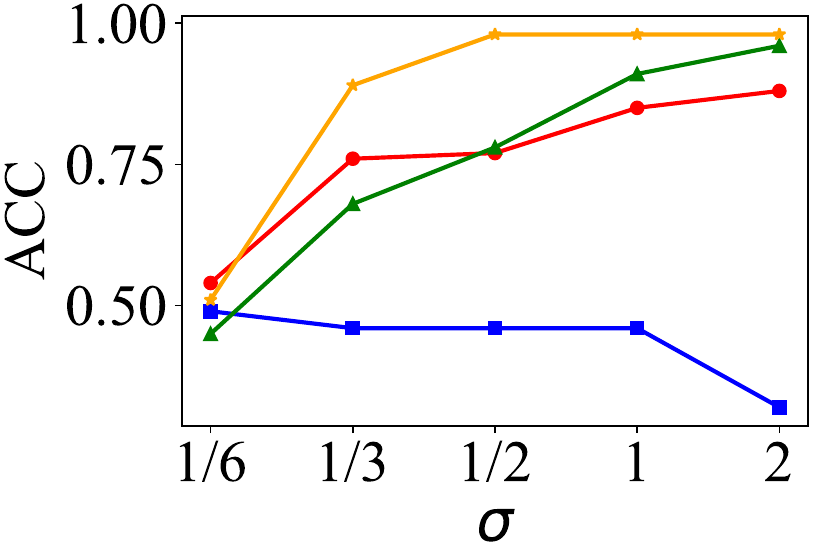}}
    % \subfloat[CNN6]{\includegraphics[width=0.49\linewidth]{figures/modeldnn_sigma_setting.pdf}}
  \end{minipage}

  \caption{Detection accuracies under different $\sigma$ of backdoor probes. The x-axis coordinate value represents multiples of $|\mathcal{S}^{\text{noise}}|^{\max}$.}
    \label{fig:sigma}
    % \vspace{-3mm}
\end{figure}

For architectures using Gaussian probes, we further study the influence of the variance parameter $\sigma$. Fig.~\ref{fig:sigma} shows that HTell remains stable within a moderate variance range, provided that the generated probes sufficiently cover the coarse latent activation region.

% According to this two indicators, we set $\sigma=\frac{1}{2}|\mathcal{S}^{\text{noise}}|^{\max}=0.25$ for EfficientNet-B3, $\sigma=|\mathcal{S}^{\text{noise}}|^{\max}$ for CNN6 ($\sigma=6$) and SqueezeNet ($\sigma=2$), and $\sigma=2|\mathcal{S}^{\text{noise}}|^{\max}=4$ for PreactResNet18. Similarly, $\sigma$ for other models is also set in this data-free method.

\subsection{Discussion and Limitations}
% However, under the open combinations of architectures and datasets, these methods experience a decrease in detection accuracy. % Thus, there are still challenges in using fixed parameters or generating parameters under limited conditions (e.g., model-free and data-free) to achieve backdoor detection in open scenarios.

\begin{figure}
  \centering
  \begin{minipage}[c]{1\linewidth}
  \centering
      \subfloat[SSD for OD \cite{cheng2024odscan}]{\includegraphics[width=0.485\linewidth]{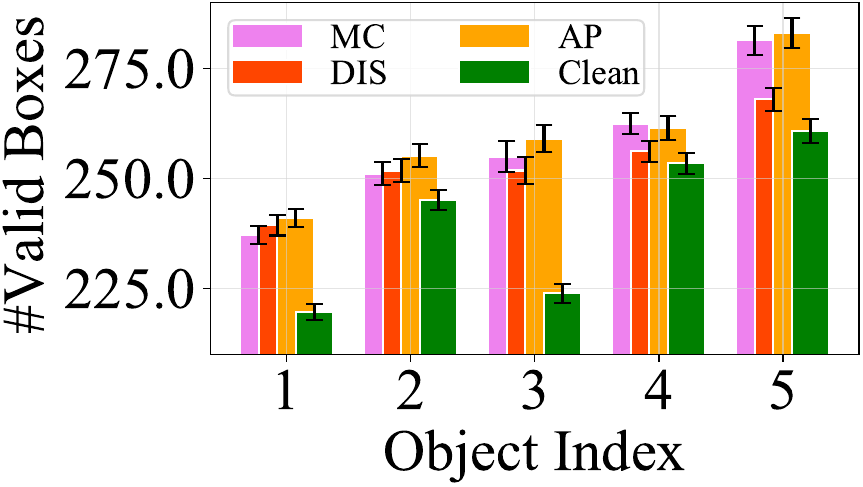}}
      \hspace{1mm}
      \subfloat[Faster-RCNN for OD \cite{cheng2024odscan}]{\includegraphics[width=0.485\linewidth]{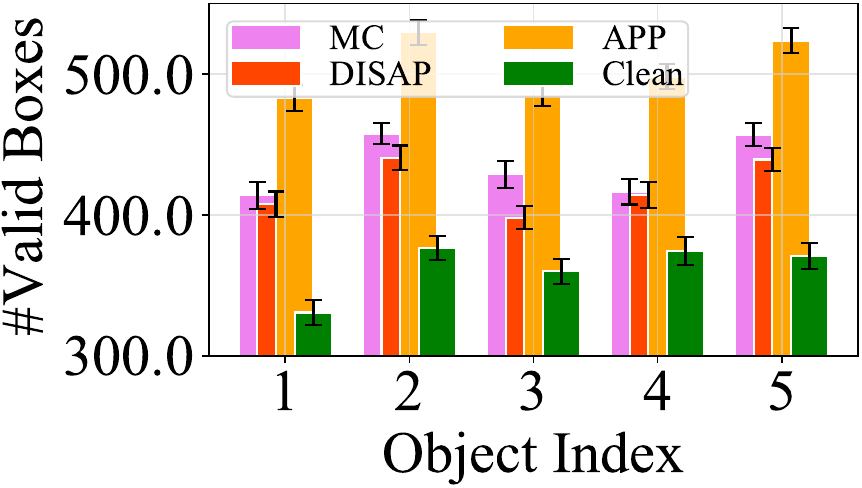}}
  \end{minipage}
  \vspace{-2mm}

  \begin{minipage}[c]{1\linewidth}
  \centering
      \subfloat[Single-Agent DRL \cite{yu2022temporal}]{\includegraphics[width=0.415\linewidth]{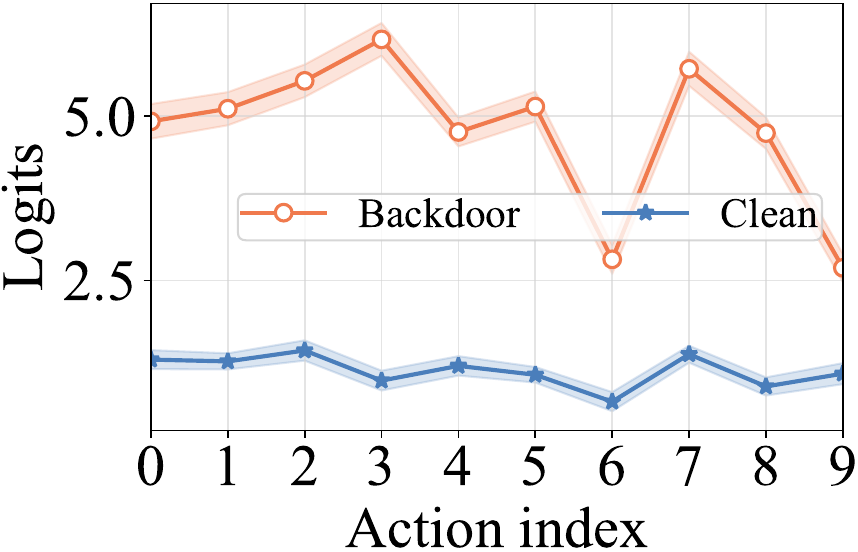}}
      \hspace{1mm}
      \subfloat[Multi-Agent DRL \cite{chen2022marnet}]{\includegraphics[width=0.56\linewidth]{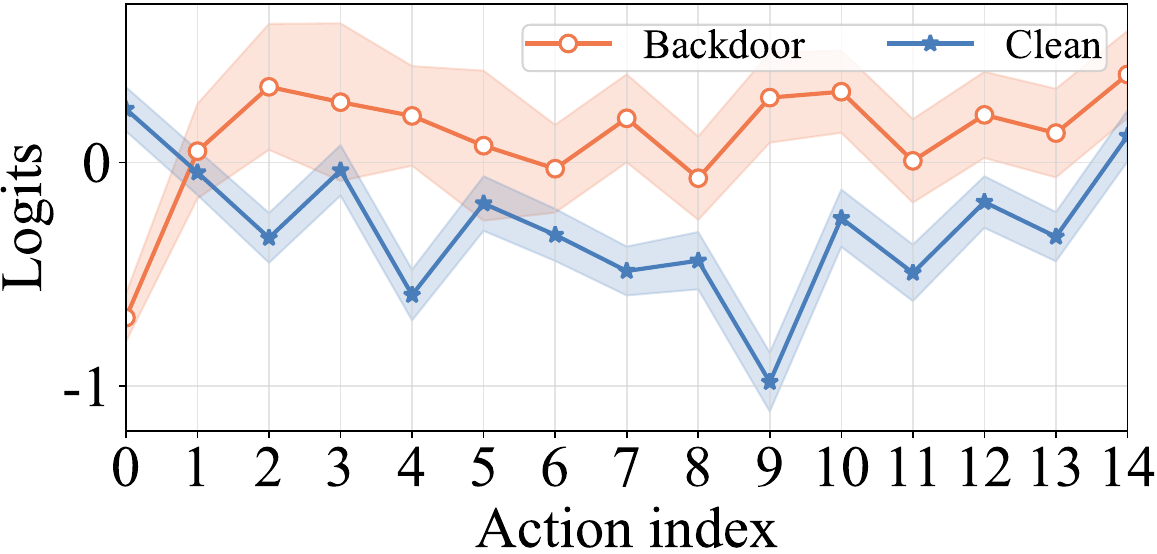}}
  \end{minipage}
  \caption{Applying HTell to object detection and sequential decision-making tasks. `MC', `AP', and `DISAP' denotes misclassification, appearingm and disappearing backdoor attacks, respectively. }
  \label{fig:applicability}
\end{figure}

We evaluated HTell on a large-scale image classification backdoor benchmark, where it demonstrated superior detection accuracy and efficiency. We further assess its applicability to other domains: object detection (OD) and deep reinforcement learning (DRL). For OD, we follow \cite{cheng2024odscan} and evaluate benign and backdoored SSD and Faster-RCNN models on the Synthesis dataset under three backdoor attacks: misclassification, appearing, and disappearing. For DRL, we train backdoored policies using attacks targeting both single-\cite{yu2022temporal} and multi-agent \cite{chen2022marnet} settings. Fig.~\ref{fig:applicability} illustrates the response divergence between benign and backdoored model heads when fed random noise. In OD, we compute valid box counts from classification logits and bounding-box regressions, comparing distributions across object indices. In DRL, we treat the final linear layer as the head and compare output logits across action indices. Clear response gaps are observed between clean and backdoored models, suggesting that head probing may extend beyond image classification. However, adapting HTell to other domain tasks requires task-specific probing points and response statistics, which we leave to future work.

Furthermore, it still has several other limitations. \textbf{First}, HTell involves two hyperparameters ($\sigma$ and $\tau$) that need to be configured.
Although HTell requires fewer trained models to determine these hyperparameters compared to existing methods \cite{fu2023freeeagle, zhang2025barbie}, this process can still be challenging for new network architectures. Existing methods either employ fixed parameters for specific fixed dataset-architecture pairs \cite{fields2021trojan, wang2024mm} or utilize more models to obtain parameters \cite{zhang2025barbie}.  However, in open scenarios with diverse combinations of architectures and datasets, these methods tend to suffer from reduced detection accuracy. Thus, achieving effective backdoor detection in open settings remains challenging, whether using fixed parameters or generating parameters under constrained conditions (e.g., model-free or data-free).
\textbf{Second}, if an adversary becomes aware of our detection mechanism, they can easily evade HTell by freezing the head parameters. This adaptive attack is effective with less performance degradation to defend against not only HTell but also many existing detectors \cite{fields2021trojan, zhou2024data}. To counter this, we would need to adaptively relocate the probes to the backbone. \textbf{Third}, HTell currently focuses on image classification tasks. Models for other non-classification tasks (e.g., object detection, semantic segmentation) typically lack a head component. Therefore, for these tasks, new probing points would need to be identified.
\textbf{Fourth}, we only studied backdoor detection without backdoor mitigation strategies. Previous studies \cite{wang2022rethinking, zhang2024exploring} have shown that neuron activations corresponding to Trojan behavior are orthogonal to others in the latent space. Given valid backdoor probes, it might be possible to reverse the activation direction of $\delta$ and apply regularization along this direction to suppress backdoor activations. We leave this extension for future work.

% we may be able to reverse the activation direction of $\delta$ and introduce regularization on this direction to suppress backdoor activations. We leave this extension as future work.

% SSL is more robust to data imbalance \cite{liu2022self}.

% A potential backdoor mitigation method is that since we have no knowledge of backdoor triggers, we may be able to utilize the abnormal indicators in the last layer to reverse the activation direction of $\|\delta\|$ (see Sec. \ref{sec:key}) via singular value decomposition on $\mathbf{W}$, similar to \cite{phan2024clean}. Considering the orthogonality \cite{zhang2024exploring}, we can perform suppression for any activations in the latent space on this trigger direction. In addition, we can also use this direction to project the exception weights and biases, and correct the last layer of parameters through parameter fine-tuning. We leave this extension as future work.

\section{Conclusion}
In this paper, we propose HTell, a novel, fast, and data-free backdoor detection framework. We identified a key insight that backdoor implantation can introduce a larger and abnormal decision boundary at the backdoor target class. This decision boundary enables the model's head to classify random latent probes as the target class. Based on this insight, we designed a simple but effective backdoor detection mechanism by generating noise in the latent space following a model-specific random distribution and querying the classifier to analyze the output logits to identify backdoors. We evaluated HTell with a large-scale backdoor detection benchmark that contains over 6,000 backdoor models involving different datasets, model architectures, backdoor triggers, and attack injection methods. Our experimental results demonstrate that compared with existing methods, HTell achieves superior performance in terms of detection accuracy and time cost across.

% \appendix
% \section*{Ethical Considerations}

%
% \section*{Open Science}
%

\bibliographystyle{IEEEtran}
\bibliography{IEEEabrv, ref}

% \appendix %% CCS: DO NOT REMOVE
%
% \input{sections/7.appendix}

\end{document}